\documentclass[english,12pt]{article}
\usepackage[T1]{fontenc}
\usepackage[utf8]{inputenc}
\usepackage{geometry}
\usepackage{epigraph}
\usepackage{amsmath}
\usepackage[plain]{fancyref}
\usepackage{microtype}
\usepackage{amsthm}
\usepackage{multirow}
\usepackage{array}
\usepackage{tabularx}
\usepackage{pdflscape}
\usepackage{amssymb}
\usepackage{afterpage}
\usepackage{longtable}
\usepackage{placeins}
\usepackage{lscape}
\usepackage{bbm, dsfont}
\usepackage{float}
\usepackage{setspace}
\usepackage{lineno}
\usepackage{booktabs}
\usepackage{subcaption}
\usepackage{afterpage} % Add this in your preamble

\usepackage{natbib}
\usepackage{graphicx}
\usepackage{stackrel}
\usepackage{mathpazo}

\usepackage[citecolor= black,
			colorlinks=true,
		    linkcolor = black,
            urlcolor  = black,
            pdfpagelabels]{hyperref}

\usepackage{tikz}
\usetikzlibrary{er,positioning}
\usetikzlibrary{calc,trees,positioning,arrows,chains,shapes.geometric,%
    decorations.pathreplacing,decorations.pathmorphing,shapes,%
    matrix,shapes.symbols}

\tikzstyle{bag} = [align=center]
\tikzset{
  decoration={border},
  tuborg/.style={decorate},
  tubnode/.style={midway, right=2pt}
}

\usepackage{hyperref}
  
\usepackage{xcolor}
  
\hypersetup{
 	colorlinks= true,
 	citecolor= blue,
 	linkcolor= blue,
 	filecolor= blue,      
 	urlcolor= teal,
 }
\newcommand{\emdatcount}{3,769 }
\newcommand{\sourcecount}{466 }
\newcommand{\countrycount}{123 }
\newcommand{\fullsample}{167,368,560}

\makeatletter
\theoremstyle{plain}

\theoremstyle{plain}

\usepackage{babel}

{
\setlength{\parindent}{0pt}
}

\newcommand\startappendixtables{%
    \makeatletter
       \setcounter{table}{0}
   \setcounter{table}{0}\global\def\thetable{A\arabic{table}}%
    \makeatother}
 
 \newcommand\startappendixfigures{%
    \makeatletter
       \setcounter{figure}{0}
   \setcounter{figure}{0}\global\def\thefigure{A\arabic{figure}}%
    \makeatother}

\makeatother

\usepackage{babel}
\providecommand{\lemmaname}{Lemma}
\providecommand{\theoremname}{Theorem}
\title{Social and Genetic Ties Drive Skewed Cross-Border Media Coverage of Disasters\footnote{We thank Marco Verile and Leonida Della Rocca.  The authors declare no competing interests.}}
\author{Thiemo Fetzer \and Prashant Garg \footnote{Garg is based at Imperial College London. Fetzer is based at University of Warwick \& Bonn and affiliated with CEPR, CAGE, NIESR, ECONtribute, Grantham Institute. Fetzer acknowledges funding by the Leverhulme Prize in Economics, a European Research Council Starting Grant (ERC, MEGEO, 101042703), and Deutsche Forschungsgemeinschaft (DFG, EXC 2126/1 -- 390838866).}}

\date{\today }

\begin{document}
% \linenumbers
\maketitle

%\author{Prashant Garg\qquad{}Thiemo Fetzer}
%\date{\today}

\begin{abstract}
Climate change is increasing the frequency and severity of natural disasters worldwide. Media coverage of these events may be vital to generate empathy and mobilize global populations to address the common threat posed by climate change. Using a dataset of \sourcecount news sources from \countrycount countries, covering 135 million news articles since 2016, we apply an event study framework to measure cross-border media activity following natural disasters. Our results shows that while media attention rises after disasters, it is heavily skewed towards certain events, notably earthquakes, accidents, and wildfires. In contrast, climatologically salient events such as floods, droughts, or extreme temperatures receive less coverage. This cross-border disaster reporting is strongly related to the number of deaths associated with the event, especially when the affected populations share strong social ties or genetic similarities with those in the reporting country. Achieving more balanced media coverage across different types of natural disasters may be essential to counteract skewed perceptions. Further, fostering closer social connections between countries may enhance empathy and mobilize the resources necessary to confront the global threat of climate change. 
\end{abstract}

%three main messages
% media reporting strongly skewed towards certain natural disaster types
% this skew is a systematic pattern between countries, not just within -- its a common across media system (attention sells) --> bias 
% deaths are an important feature explaining variation in estimated reporting increase again systematically across countries
% deaths and where the disaster occurred

% work out what is the "attention formula"
%I want to figure out or find out for WHICH countries is the location of disaster a substitute to the dyadic variables or other variables in achieving good "fit"

%this way one could tease out substitutability between location anti-Africa bias, with more nuanced measures

%to do:

% - dyadic measures level effect 
% - dyadic measures interaction with death
% - Shapley values dyadic weighted by event deaths importance by country
% - Shapley values disaster deaths only weighted by event deaths importance by country
% - Shapley values both variables weighted by event deaths importance by country
% --> compare the values to look at the information gain e.g. we hope to see 

\section{Introduction}
Climate change is a pressing global risk that affects all of humanity, both directly and indirectly, often through complex and unforeseen mechanisms \citep{IPCC2018, Stern2007Economics}. As climate change intensifies, natural disasters become more frequent and severe, posing significant threats to societies worldwide \citep{Field2012Managing, Coumou2012Decade, IPCC2021Climate}. Yet, traditional media systems frequently appear “blind” to climate-induced disasters that are less visually dramatic, such as floods, storms, and extreme temperatures, focusing instead on events like earthquakes and wildfires \citep{Boykoff2009WeSpeakForTheTrees, Doulton2009TenYears}.
% This is particular vital, since the severity of climate change and its consequences may manifest itself differently across different countries. 

Accurate and comprehensive media coverage is essential for raising public awareness of climate change and encouraging global cooperation \citep{Boykoff2007Climate, Painter2013Climate, Schafer2015Climate}. Media reporting helps shape perceptions of climate-related risks and can influence public support for mitigation and adaptation policies \citep{Pralle2009AgendaSetting, McCombs2004SettingAgenda, Gavin2009Climate}. However, the media often devotes disproportionate attention to certain events or regions, potentially skewing public perceptions \citep{Soroka2019BadNews, Zhuravskaya2020PoliticalEffects, Prat2011MediaInfluence}.

Global action on climate change requires coordination and cooperation across borders \citep{Hulme2009WhyWeDisagree, Beck2009WorldRisk}. Media coverage not only shapes domestic opinions but also influences global policy debates and social norms, making transnational perspectives essential for fully understanding the interplay between climate issues and public discourse \citep{schafer2018media, entman2009projections}. Empathy—encouraged by comprehensive media reporting that clearly links climate-related disasters and their human consequences to their underlying causes—can play a key role in building global solidarity \citep{Batson2011Altruism, Hoffman2001Empathy, deWaal2009AgeOfEmpathy}. Empathy is considered a pinnacle of human moral development and helps motivate prosocial behavior, moral responsibility, and cooperative actions \citep{Hoffman2001Empathy, Tomasello2019BecomingHuman, Decety2014ComplexRelation, Goleman1995Emotional}. Media reporting that potentially enhances empathy towards those affected by climate change-induced disasters could be important for fostering global cooperation.

% Understanding cross-border media coverage is particularly important, as the consequences of climate change are global and interconnected \citep{Stevenson2015Overcoming, Hulme2009WhyWeDisagree, Beck2009WorldRisk}. Reporting on natural disasters occurring elsewhere can foster empathy and a sense of global community, which are essential for collective action \citep{Batson2011Altruism, Hoffman2001Empathy, deWaal2009AgeOfEmpathy}. Empathy, a complex cognitive and emotional capacity, is crucial in motivating prosocial behavior and moral responsibility \citep{Decety2014ComplexRelation, Goleman1995Emotional, Tomasello2019BecomingHuman}. It plays a central role in moral development and is a predictor of prosocial behavior and well-being \citep{Batson2011Altruism, Tomasello2019BecomingHuman}. The media serves as a critical vector for evoking empathy by making distant suffering more immediate and emotionally resonant \citep{Kearney2013CrossCulturalEmpathy, Decety2010Empathy}.

Despite the increasing prevalence of climate change-induced natural disasters, and the importance of empathy and global cooperation, there is limited research on cross-border media reporting of these events and potential biases therein. Most empirical work has focused on individual countries, particularly the United States, providing a deep understanding of the political economy of media, economics, and politics \citep{Gentzkow2006MediaBias, DellaVigna2007FoxNews, Bursztyn2020Misinformation}. However, transnational perspectives are essential to address global challenges like climate change.

Despite the increasing prevalence of climate change-induced disasters and the importance of empathy in promoting global cooperation, relatively few studies have examined cross-border media reporting on these events. Most existing work focuses on single countries, often the United States, offering insights into the political economy of media and politics \citep{Gentzkow2006MediaBias, DellaVigna2007FoxNews, Bursztyn2020Misinformation}. Yet, addressing global challenges like climate change demands a broader, transnational perspective.

In this study, we analyze whether media sources linked to one country report more on another country following a natural disaster. We use a unique open-source intelligence data resource, the Europe Media Monitor (EMM) \citep{Atkinson2009NearRealTime, Steinberger2009JRC,jacquet2020covid}, developed by the European Commission’s Joint Research Centre. The EMM compiles data from billions of publicly available online sources, employing natural language processing algorithms to extract key information—such as article content and event locations—from this vast media universe.

Combining the EMM dataset with the Emergency Events Database (EM-DAT) database of global natural disasters \citep{EMDAT2021}, we construct a dyadic panel capturing cross-country media reporting before and after each of \emdatcount disasters. Our dataset includes \sourcecount sources from \countrycount countries, amounting to over 135 million articles indexed since 2016. 

We estimate whether media outlets in one country systematically increase coverage of another country following a disaster, using a co-occurrence approach of article counts mentioning affected countries. With these estimates, we study how coverage varies by disaster type and by the severity, measured in fatalities, of the events. We then explore to what extent ties between countries correlate with the intensity and patterns of coverage. This analysis quantifies the extent of how existing biases in newsworthiness and global connectedness skew the portrayal of natural disasters.

%\textbf{Note to add on dysfunction and spread of misinformation into media ecosystems during the COVID19 pandemic} The Europe Media Monitor provided a first open access dataset early in the pandemic \cite{jacquet2020covid}.

%Our research aims to identify systematic patterns and biases in media reporting on natural disasters, particularly those linked to climate change, and assess how these may affect public awareness and global cooperation. We focus specifically on cross-border media coverage to understand whether natural disasters in one country are reported on in media associated with other countries. This may increase our understanding of the extent to which biases or skews in what media may consider newsworthy may affect perceptions of climate change and its consequences.  

Our findings suggest that media coverage is heavily skewed towards events that are visually salient, such as earthquakes and wildfires. In contrast, climatologically significant disasters like floods, droughts, storms, and extreme temperatures receive far less attention. This pattern appears to be remarkably consistent across countries, suggesting that it is a general feature of global media reporting. This aligns with findings that visually compelling imagery is crucial in eliciting public engagement and shaping perceptions of climate issues \citep{o2014climate}
% This skew in reporting may contribute to shaping public perceptions of climate change. Although we hypothesize that countries with higher media coverage of earthquakes and wildfires could have lower public awareness of climate change, this relationship remains to be demonstrated directly.

The number of deaths caused by a natural disaster appears to be one of the strongest predictors of the average increase in media coverage. The media in most countries exhibit a noticeable \emph{death gradient}, where the degree of reporting increases with the number of fatalities. However, the influence of death count on media coverage varies significantly between countries, indicating that this relationship is not uniform.

When decomposing what drives differences in reporting patterns between countries, we find that historical measures of country connectedness, on their own, do not adequately explain these variations. Instead, a combination of natural disaster characteristics and measures of country-to-country connections plays a key role. Specifically, the explanatory power of the number of disaster deaths is significantly enhanced when considering the strength of social ties between countries. Genetic similarity also appears to be influential, possibly reflecting deeply rooted in-group preferences shaped by evolutionary and sociocultural factors \citep{simpson2015beyond, choi2007coevolution}. Such parochial altruism and the moral significance of in-group membership may make certain disasters more ‘relatable’ and thus more newsworthy. This suggests that human connections, whether cultural, social, or historical, play an important role in fostering empathy and shaping how the media respond to disasters in other countries.

%average dyadic counts across all observation is highly skewed Mean 0.946 for all observations, and 0.579 for observations that focus on other countries (i.e. where the country pair is not US US, but, say US - Other country).
% after the effect, add something like (effect size: \textbf{0.25}, 95\% CI: \textbf{0.10--0.40})

This study contributes to a deeper understanding of how skewed media reporting may shape perceptions of climate-related risks and highlights the need for more balanced media reporting to foster global awareness, empathy, and cooperation in addressing climate change. By shedding light on the systematic patterns and biases in media reporting, we hope to inform more effective strategies for addressing the global challenge of climate change.

\section{Results}

\subsection{Media coverage following a natural disaster}
Figure \ref{fig:averagereportingincrease}(a) displays the average increase in the volume of news articles attributed to a country after that country experiences a natural disaster. This figure focuses on the type of natural disaster and shows the average estimated increase in reporting within three days following the event. We find that, after a natural disaster occurs, media reporting by countries not directly affected rises unevenly, depending on the type of disaster. The increase is most pronounced for earthquakes, technological disasters (such as, for example transportation accidents), and wildfires. Specifically, the average increase in media reporting on a country affected by an earthquake, technological accident, wildfire, or storm is 7.4\% (b = 7.38\%, 95\% CI [6.49,13.03], p < 0.001), 2.87\% (b = 2.87\%, 95\% CI [2.54,4.93], p < 0.001), 2.08\% (b = 2.08\%, 95\% CI [0.42,3.92], p < 0.05), and 1.33\% (b =1.33, 95\% CI [0.79,1.79], p <0.01), respectively. In contrast, we do not detect a robust and systematic increase in reporting when a country is affected by floods, droughts, or extreme temperatures.

%Panel B explores the extent to which the pattern hat is detected across the different natural disasters 
To check the robustness of these findings, Appendix Figure \ref{fig:averagereportingincrease_flexible} explores a more flexible time window of -7 to +7 days relative to the start date of the natural disaster. This wider window accounts for potential anticipatory reporting (e.g., for floods) and confirms that the main pattern remains intact. Thus, focusing on the three-day window after the disaster start date provides a suitable estimate of pooled reporting effects.

To verify that the skewed reporting pattern by disaster type is consistent across reporting countries, Figure \ref{fig:averagereportingincrease}(b) plots the country-specific estimated increases for each disaster type (vertical axis) against the country-specific estimated increase for earthquakes (horizontal axis). Each point represents a reporting country, and the fitted regression line is shown without an intercept for visual ease. Although there is considerable heterogeneity, the overall pattern—more pronounced reporting on earthquakes, technological disasters, and wildfires—is systematic across countries.

\subsection{Fatalities are an important driver of media reporting}
\begin{landscape}
\begin{figure}[H]
    \centering
    \caption{Media reporting increase strongly informed by the number of deaths}
    \vspace{0.2cm}

                \includegraphics[scale=0.55]{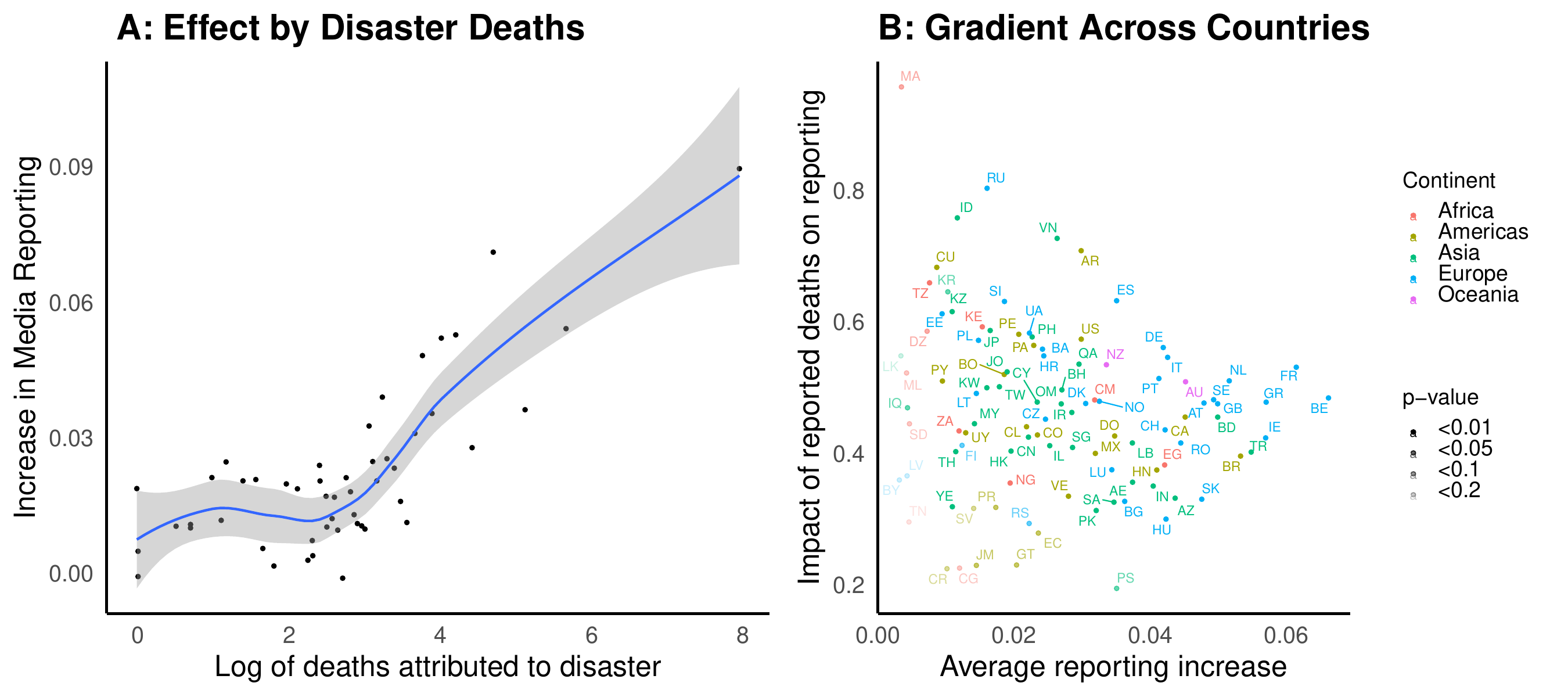}

%    \begin{tabular}{cc}
%        \begin{minipage}{0.7\textwidth}
%            (a) Reporting is strongly but non-linearly increasing in the number of deaths \\
%            \includegraphics[width=\textwidth]{figures/estimate-total-deaths-logp1-dyadic_counts-3.pdf}
%        \end{minipage}
%        &
%        \begin{minipage}{0.78\textwidth}
%            (b) Death gradient strength relative to average reporting varies notably across countries. \\
%            \includegraphics[width=\textwidth]{figures/res-estimate-country-death-gradient-ratio-dependent-variable-dyadic_counts-logp1-3.pdf}
%        \end{minipage} \\
%    \end{tabular}
    \captionsetup{singlelinecheck=off,font=scriptsize}
    \caption*{\textbf{Note}: Figure displays the estimated increase in dyadic media coverage across \emdatcount natural disasters that occurred since 2016 by natural disaster type. We construct an event-level panel dataset that captures dyadic news coverage across the \sourcecount sources in \countrycount countries. For each natural disaster occurring in a country $i$, we construct a 7-day window around the disaster start and end dates. For this event dataset, we attach a dyadic daily dataset measuring the number of media articles of a news source $s$ in country $j(s)$ that mentions or refers to news covering any other country. Throughout, we remove the distorting effects of any other time-varying shock and the average level differences in the propensity of a news source to cover any country. Panel A highlights that the estimated increase in news reporting, across disasters, is strongly and non-linearly increasing in the number of deaths. Panel B highlights that this death gradient varies across countries. In Panel B, the x-axis captures the average country-specific reporting increase, while the y-axis is the ratio of the estimated country-specific death gradient divided by the value of the x-axis to see the two in relations. }
    \label{fig:deathgradient}
\end{figure}
    \end{landscape}

Figure \ref{fig:deathgradient} highlights another salient factor influencing the increase in media reporting after a natural disaster: the number of fatalities. Panel (a) of Figure \ref{fig:deathgradient} presents the results of a non-linear regression, showing that the average increase in media coverage rises substantially when a natural disaster causes more than 20 deaths. This suggests threshold effects may influence editorial decisions on whether and how much to report on foreign disasters.

In Figure \ref{fig:deathgradient}(b), we examine how this “death gradient” in media reporting varies across different reporting countries. The horizontal axis plots, for each reporting country, the average estimated increase in reporting following a natural disaster, regardless of the disaster’s characteristics. The vertical axis shows the country-specific death gradient, obtained from a linear regression estimated separately for each reporting country. This death gradient estimate is normalized by the country’s average reporting increase (shown on the horizontal axis), allowing us to relate the relative strength of the death gradient to a country’s baseline reporting level.

The results indicate that the positive impact of fatalities on media coverage is universal. However, countries that have lower baseline reporting increases show a more pronounced death gradient. In other words, the number of deaths drives reporting to a greater degree for countries that, on average, devote less coverage to foreign disasters. This finding suggests that, for media sources attached to these countries, the severity of a disaster (as measured by fatalities) is a particularly important cue for deciding whether to cover it.

Naturally, this result does not imply that the public necessarily pays attention to the detected statistical patterns in reporting. Rather, it indicates that in certain countries, the likelihood of reporting on a foreign disaster is heavily skewed by whether the disaster has caused a substantial number of immediate casualties.

\subsection{Disaster characteristics and bilateral country connections jointly explain notable reporting differences}
So far, our analysis has focused on how disaster characteristics—specifically type and fatalities—explain variations in reporting patterns across countries. We have found substantial heterogeneity: while disaster type and fatalities significantly influence media coverage, their impact is not uniform across different reporting countries.

% The analysis suggests complex interaction effects and highlights large heterogeneity in the estimated reporting increases $\hat{\beta}_{k(j),i}$.
This complexity suggests that additional factors, such as the nature of connections between countries, may interact with disaster attributes. Appendix Figure \ref{fig:dyadic_estimate_univariate} presents univariate relationships showing that various measures of country connectedness—including historical ties, cultural or linguistic similarities, values-based alignments, social connections, and genetic similarities—are generally positively associated with increased reporting on disasters in the other country. For example, country-pairs with stronger social ties or similar languages or religious compositions tend to report more on each other’s disasters. However, these associations are relatively small, indicating that connectedness measures alone do not account for much of the observed variation in reporting patterns.

To better understand these interactions and the underlying heterogeneity, we employ a random forest regression to decompose the variation in the estimated reporting increases $\hat{\beta}_{k(j),i}$. The random forest approach is well-suited for this task because it can capture non-linearities and interaction effects. In learning the random forest, we focus on the set of dyadic features that are consistently available—this choice excludes genetic dissimilarity data, which is only available for a smaller subset of country pairs.

Panel A of Figure \ref{fig:country-connectedness} shows how well the random forest models perform using three different sets of features: (1) the dyadic measures of country connectedness, (2) the characteristics of the natural disasters, and (3) a combination of both sets. When we use only dyadic connectedness measures, the best cross-validated random forest explains about 0.5\% of the variation in reporting increases. Using only disaster characteristics—namely, disaster type, number of fatalities, and disaster duration—improves the explained variation to around 12\%. Combining both sets of predictors raises the $R^2$ to about 19\%.  Although a large portion of the variation remains unexplained, this marked improvement suggests that country-connectedness measures interact strongly with disaster characteristics to shape cross-border reporting patterns. Appendix Figure \ref{fig:robustnessrandomforests-variable-transformations} confirms that this pattern is not an artifact of the chosen functional form for estimating $\hat{\beta}_{k(j),i}$, nor is it driven by the particular dependent variable specification—dyadic occurrence counts—used in the analysis.

Panel B of Figure \ref{fig:country-connectedness} compares variable importance measures for the disaster characteristics in models that include and exclude the dyadic connectedness features. Here, we report scaled permutation importance scores to ensure comparability. The results show that incorporating country connectedness increases the relative importance of the number of fatalities as a predictive factor. In other words, the explanatory power of fatalities in shaping media coverage appears to be enhanced when we account for how connected the countries are. This suggests that the perceived “newsworthiness” of a natural disaster, as proxied by fatalities, interacts with country-level linkages to influence media reporting decisions.

Panel C in Figure \ref{fig:country-connectedness} explores which specific country-connectedness features interact most strongly with the number of fatalities. To do this, we estimate a series of linear models, each introducing more stringent sets of fixed effects, and convert dyadic features into z-scores to facilitate comparisons. Two variables stand out: the degree of social connections between countries and their genetic similarity. Although genetic similarity data are available for fewer pairs of countries, its importance is still notable. By contrast, factors such as colonial ties, trade linkages, or religious or linguistic similarity of the populations appear to have much weaker effects on cross-border media reporting.

Overall, these findings indicate that social connections between the peoples of different countries are a primary factor enhancing media coverage of natural disasters abroad, especially as the number of fatalities increases.

\paragraph{Visualising these patterns} 
To make our findings and their implications more salient, we develop an interactive visualization that illustrates cross-country differences in how media report on natural disasters.\footnote{This visualization, along with other plots that facilitate data exploration, is available at \url{https://www.trfetzer.com/megeo-dyadic-news/}.} Specifically, we use the final random forest model to generate out-of-sample predictions, $\widehat{\beta}_{i,j}$, for a suite of hypothetical yet common natural disasters occurring in selected locations. Each hypothetical disaster is assigned varying numbers of fatalities to capture different severity levels.

This procedure enables us to systematically plot, for a given \emph{reporting country}, how the estimated reporting increase $\widehat{\beta}_{i,j}$ changes with the disaster’s death toll. Figure~\ref{fig:counterfactual-illustration} illustrates this approach by focusing on two reporting countries—Germany and India—and three hypothetical disaster locations: Bangladesh, Italy, and Mexico. In this example, each scenario is a storm with a varying number of fatalities.

The fitted values allow us to make like-for-like comparisons across countries and scenarios. For instance, from the perspective of the German media, a storm causing 100 fatalities in Bangladesh elicits roughly the same relative reporting increase as a storm with about 43 fatalities in Italy or 62 in Mexico. Conversely, for Indian media, a storm with 45 fatalities in Bangladesh yields the same approximate coverage as a storm with 100 fatalities in Mexico or Italy. These simulations highlight a core observation: although the absolute severity of a disaster matters, factors like cultural proximity, social connectedness, and genetic similarity significantly modulate how foreign media perceive and prioritize an event.

By conducting such counterfactual comparisons, we show that two disasters with the same death toll can generate markedly different levels of coverage, depending on the countries involved and the specific audiences. Hence, what may appear “close” or “relevant” to one set of media outlets—and, by extension, their audiences—can be strikingly different for another. This insight aligns with our broader finding that media coverage is not solely driven by the raw number of casualties but also by deeper social, cultural, and evolutionary ties between populations.

\section{Discussion}
Our study provides evidence that cross-border media reporting after natural disasters is heavily skewed toward certain event types—particularly earthquakes and wildfires—while underreporting other climate-relevant disasters such as floods, droughts, storms, and extreme temperatures. This imbalance in coverage may have far-reaching implications for public understanding of climate change, the growth of global empathy, and the development of international cooperation to address this urgent challenge.

Climate negotiations are often fraught with tensions, exacerbated by continued efforts to cast doubt on the scientific consensus. Media reporting may play a crucial role in fostering public awareness of both, the scientific consensus as well as the consequences of climate change.  Rising geopolitical tensions, which risk fragmenting the global information sphere, may provide avenues for countries less concerned about the consequences of climate change to suppress reporting on adverse climate impacts, such as natural disasters. This suppression of information risks undermining popular support for climate action at a critical time.

This concern is heightened by the key role that elite messaging and media narratives play in influencing public perceptions of climate change. For example, \cite{merkley2020party} highlights how partisan media can foster skepticism. Social norms and elite endorsements remain critical for garnering public support for mitigation policies. \cite{rinscheid2020public} stresses that descriptive norms and elite cues can shape climate policy approval, emphasizing the need for careful, strategic communication. In an era defined by fragmented media and geopolitical discord, effectively using these tools is vital to counter skepticism and build momentum for global climate cooperation.

Climate change presents risks not just through direct impacts, but also via complex, cascading effects \citep{IPCC2018, Beck2009WorldRisk}. The increasing prevalence of climate-related disasters underscores the urgency of global action \citep{Field2012Managing, Coumou2012Decade}. However, traditional media often appears “blind” to many climate-induced natural disasters, especially those that lack dramatic visuals or immediate effects \citep{Boykoff2009WeSpeakForTheTrees, Doulton2009TenYears}. This underreporting may limit public appreciation of the full, multifaceted nature of climate risks \citep{Schmidt2013MediaAttention, Painter2013Climate, Moser2010Communicating}.\footnote{How climate risks and responsibilities are communicated can strongly influence both public engagement and policy outcomes \citep{Dryzek2011Oxford, stevenson2012discursive}. Empirical work on climate change communication suggests that more balanced and inclusive reporting can foster more robust public debate and enhance global solidarity \citep{moser2016reflections}.}

Our findings also show that increases in reporting are strongly skewed toward disasters that cause significant fatalities, implying potential threshold effects in drawing media attention. This relationship, however, varies considerably across countries. Research on the link between media coverage and public awareness, especially regarding global challenges like climate change, is extensive \citep{Pralle2009AgendaSetting, McCombs2004SettingAgenda}. Media bias—where certain events are overemphasized—has a substantial influence on public understanding of critical issues \citep{Zhuravskaya2020PoliticalEffects, Prat2011MediaInfluence, Lacy2001EffectsOfContent}. Such editorial decisions align with long-established principles of newsworthiness, where sudden, dramatic events are more likely to receive extensive foreign coverage \citep{galtung1965structure}. Similar patterns have been documented more broadly in journalism studies, where certain event attributes systematically increase the likelihood of extensive reporting \citep{wahl2009handbook}. The adage “if it bleeds, it leads” captures the media’s tendency to focus on dramatic events \citep{Soroka2019BadNews, Harcup2001NewsValues}, which may distort perceptions of climate change by emphasizing sudden catastrophes over gradual, equally consequential changes.

A significant factor shaping these reporting biases is the interaction between disaster characteristics and country connectedness. Social ties, as measured by social connectedness, are particularly influential: media outlets devote more coverage to disasters in countries that share closer social bonds, even if the number of fatalities is the same. Genetic similarity between populations plays another central role. We can not distinguish whether this is due to genetic similarity being associated with stronger social ties and the associated improved information flows or whether it may capture a structural factor that may increase relatedness.  

Empathy is crucial for motivating prosocial behavior and global cooperation, and is considered central to human moral and cognitive development \citep{Hoffman2001Empathy, Tomasello2019BecomingHuman, Decety2014ComplexRelation}.\footnote{Theories of empathy, such as Martin Hoffman's "empathic arousal" \citep{Hoffman2001Empathy}, suggest that empathy is triggered by the ability to emotionally and cognitively relate to the suffering of others.} Media coverage can evoke empathy by bringing distant suffering closer, both emotionally and cognitively \citep{Kearney2013CrossCulturalEmpathy, Decety2010Empathy}. However, when coverage disproportionately focuses on dramatic disasters, it may trigger selective empathy\footnote{Empathy alone may not guarantee constructive collective action, particularly when it is narrowly focused or selective. Research suggests that expanding the circle of moral concern and understanding group identity dynamics is essential for fostering broader empathy \citep{brewer2007importance}, while cautioning against uncritically assuming empathy will always yield beneficial outcomes \citep{bloom2017against}}, overshadowing victims of less visually striking events like floods or droughts. Similarly, skewed coverage toward disasters with higher death tolls or in countries deemed more relatable may distort the global picture. Such biases in empathy distribution could affect public engagement with climate issues and shape policy preferences \citep{Batson2011Altruism, Goleman1995Emotional}.

% Theories of empathy, such as Martin Hoffman's "empathic arousal" \citep{Hoffman2001Empathy}, suggest that empathy is triggered by the ability to emotionally and cognitively relate to the suffering of others. Visual imagery and real-time reporting of events such as earthquakes and wildfires may more effectively evoke empathic responses due to their dramatic nature. In contrast, events like droughts or heatwaves, which are less visually immediate and unfold over longer periods, may not evoke the same emotional reaction. Similarly, natural disasters, irrespective of type, may produce different levels of fatalities and it may be that countries' that seem closer culturally or due to established social ties may render certain events more relatable to an audience in a scarce attention economy. This would be a demand-side explanation. Alternatively, it could be that social connectedness increases the flow of information and, through this channel, increases the reporting. Either way, we observe that this may distort the representation of verifiable events. This bias in how empathy is distributed across different types of natural disasters, or casualties in different contexts may influence public engagement and shape policy preferences related to climate change \citep{Batson2011Altruism, Goleman1995Emotional}.

Media-induced empathy—or its absence— may have wide-reaching implications for global climate efforts. Natural disasters are tangible manifestations of climate change, and accurate, balanced reporting on these events can foster empathy and solidarity \citep{deWaal2009AgeOfEmpathy}. If media outlets focus narrowly on certain types of disasters, they risk skewing public perceptions of climate change’s risks and reducing empathy for underreported events, potentially weakening the impetus for collective action. Since empathy drives prosocial behavior and collective action, correcting these biases in coverage is essential for building the global cooperation needed to address climate change \citep{Hulme2009WhyWeDisagree, Beck2009WorldRisk}.

Empathy in response to media coverage is influenced not only by the type of disaster but also by the cultural, social, and ancestral proximity between countries. Studies in political economy show how ancestral links and transnational relationships shape economic and political ties \citep{Burchardi2019Migrant, Bahar2018Migration}. Thus, media reporting on disasters may serve as a conduit through which “soft power” is either amplified or diminished \citep{Nye2004SoftPower}. Our data indicate that simple dyadic measures of connectedness alone do not fully explain these patterns. Instead, the interaction between disaster characteristics and connectedness, especially social ties and genetic similarity, significantly influences cross-border media coverage. These findings suggest that media reporting may disproportionately highlight disasters that are likely to generate empathy and a sense of relatedness, reinforcing unequal attention to certain regions or disaster types \citep{Petrova2011NewspapersPoliticalCompetition, Fan2013Ownership}.

Our study has several limitations. The EMM dataset may not uniformly cover all media sources, and EM-DAT may have incomplete records for less severe disasters or those in underrepresented regions \citep{Jones2022Evaluating, Below2009DisasterDatabase}. Further research is needed to clarify the causal mechanisms behind the associations we observe.

Future investigations should identify root causes of media bias in disaster coverage and its direct effects on public perceptions, empathy, and policy preferences. While this study provides a statistical inference approach to detect average patterns across different disasters, experimental or quasi-experimental studies could directly test how exposure to various disaster types affects climate awareness and empathy. Additionally, examining the influence of social media and other information channels could yield a more comprehensive understanding of how information spreads in the digital era \citep{Allcott2017SocialMediaFakeNews, Vosoughi2018SpreadOfMisinformation, gargfetzer2024}. For instance, the observed gradient in reporting tied to social connections may be driven by the ease of information flow rather than culturally conditioned empathy.

Despite these uncertainties, our findings highlight the need for more balanced and comprehensive media coverage of climate-related disasters to enhance global awareness, empathy, and cooperation. Media organizations have a responsibility to report on a wide range of events, emphasizing the links between climate change and natural disasters. Policymakers and educators can also strengthen public understanding of climate risks, working to counteract reporting biases \citep{Boykoff2011WhoSpeaksForClimate, Moser2007MoreBadNews}. By promoting accurate, inclusive coverage, it may be possible to deepen global solidarity and inspire the collective action required to mitigate the escalating risks of climate change.

\section{Methods}
\label{section:Methods}
\subsection{Primary Data Sources}

We draw on two primary data sources. First, we use the Europe Media Monitor (EMM), an open-source intelligence platform developed by the Joint Research Centre of the European Commission \citep{Atkinson2009NearRealTime, Steinberger2009JRC}. The EMM continuously scrapes publicly available online sources—covering text and images—from thousands of news outlets in up to 70 languages, accumulating billions of items since the early 2000s. Employing web scraping and natural language processing (NLP) algorithms, it extracts relevant information such as content and geographic references. The EMM operates under a legal mandate from the European Commission, making it a robust and comprehensive dataset for analyzing cross-country media coverage.

Second, we incorporate data from the Emergency Events Database (EM-DAT) \citep{EMDAT2021}, maintained by the Centre for Research on the Epidemiology of Disasters (CRED). EM-DAT records global natural disasters, providing details such as type, location, duration, and number of fatalities. Widely used in disaster risk research and policy \citep{Below2009DisasterDatabase}, EM-DAT offers a reliable reference for identifying and classifying disaster events since 2016.

\paragraph{Dataset Construction}
The EMM capacity has grown rapidly with the expansion of the internet from its inception with a large number of sources that have been (semi) consistently indexed and made accessible through an OpenSearch interface.  A prime challenge with media observatory work leveraging primary sources across countries is the lack of data consistency across content providers over longer duration. That is, news sources may enter- and exit a news index, rendering longitudinal comparisons as highly difficult. The EMM capacity, due to its scale and breadth, made it possible to define a consistent set of sources. For the period from 2016 to 2023 we have identified a sample of \sourcecount   sources that are attached to \countrycount countries that appear consistently indexed. We proceeded in the source selection using the following filtering criteria. A source is considered if:

\begin{itemize}
\item The source must have been consistently indexed by EMM since 2016, producing near-daily articles. This ensures stable long-term coverage.
\item The source must be a primary content provider rather than a content aggregator or platform (e.g., not Google News or MSN News).
% to be finished
% \item we attach 
% \item 
\end{itemize}

Since the EMM has extensive coverage in Europe, including local and regional outlets, we further restrict the sample to achieve a more balanced country-level representation. For each country, we include at most five sources. To determine the top sources for each country, we use web traffic and search rank data from Semrush. \footnote{For example, for the US the listing is provided on \url{https://www.semrush.com/trending-websites/us/newspapers}. We retrieve the associated rank and traffic data for each country.} For countries where fewer than five sources have available traffic data, we rank remaining sources by their total indexed articles between 2016 and 2023 and select the top ones.

Applying these filters results in a sample of \sourcecount sources associated with \countrycount countries. Each source provides a consistent time series of articles from 2016 to 2023, yielding 134 million articles and allowing cross-country comparisons unaffected by source turnover.

\paragraph{Geographic attribution} The geographic aboutness of a news source is based on the country of attribution of a source. That is, the country $i$ that is attached to a source $s$ is static. The country that a specific news item may cover or relate to is retrieved from the EMM's in house geographic parsing solution. The EMM in-house approach leverages both a combination of country-specific gazetters as well as named entity recognition to identify names of places, organisations or people in the original news articles. The EMM geographic attribution pipeline has a range of disambiguation techniques in place to handle ambiguous place names. Each indexed article is tagged with a geographic aboutness or more if one has been detected. For the purpose of geographic attribution of an article to a country $j$ the country that is attached to the most frequently occurring locations is considered as the articles geographic aboutness. 

\paragraph{Language handling} Our sample includes \sourcecount sources publishing in 44 languages. Figure \ref{fig:sources_by_language} tabulates the number of sources by language: about 22\% of sources are in English, and the top 10 languages cover 72\% of all sources. EMM is explicitly multilingual with a skew towards European languages. The indexing focuses on the title, the first paragraph of a text as this would traditionally contain the most relevant and important information around a natural disaster event. For languages that are not natively supported, the EMM machine translates the content into English.  

\paragraph{Dyadic count measure} Having mapped each source $s$ to a country $i(s)$ and attributed a geographic location $j$ to each article, we construct a measure of country-pair-level article counts as a time series. To achieve this, we use OpenSearch queries to generate a structured dataset where, for each source $s$, each date $t$, and each possible destination country $j$, we count the number of articles indexed by that source that reference country $j$ on that specific date. This approach yields a fully dyadic dataset, capturing all source-destination-country combinations over time.

\paragraph{Natural disaster keyword based measure} 
In addition to raw dyadic article counts, we refine our dependent variable using a keyword-based filter. This method relies on the EMM’s categorization engine, which employs keyword indexing developed by media analysts at the Joint Research Centre. These filters focus specifically on natural disaster-related content. While our primary analysis uses raw country-pair-specific dyadic counts to capture general attention, the keyword-based measure allows us to isolate and validate patterns specifically tied to natural disasters. 

This approach may introduce idiosyncrasies or measurement errors. However, our empirical design addresses potential concerns by centering the analysis on robust, statistically detectable reporting increases. Details of the statistical adjustments and validation steps are described below.

%\paragraph{Refinement with category filters} The pure dyadic count is a crude measure that captures the overall number of articles covering a specific destination for each source on each specific day. This can be thought of as a crude measure of \emph{foreign attention}. This is desirable as ultimately we are interested in exploring to what extent news media attention on a specific country that is affected by a natural disaster increases. Yet, for a refinement we also construct a more narrow set of news articles that we are querying. 

\subsection{Empirical Design of Event Study}
Our analysis employs an event study design centered around each natural disaster event. Let $E_{k(j),t}$ denote a disaster $k$ occurring in country $j$ at time $t$. Given that the start and end dates of natural disasters may be imprecise, we account for this with flexible time windows in our model.

The dependent variable in our analysis dataset $y_{s(i)jt}$ provides for a dyadic daily count dataset measuring the number of times a news source $s$ attached to country $i(s)$ mentions or refers to news covering any other country $j$ on day $t$ either across all articles or, in natural disaster keyword filtered articles. In our case this dyadic dataset has dimensions $\sourcecount$ news sources $\times$ 123  countries $\times 8$ years $\times 365$ days =  \fullsample.  

Since the focus of this paper is specifically on \emph{spillovers}, we remove the observations $y_{s(i)jt}$ where $s(i) = j$. That is, we exclude own country pairs to ensure our media reporting differential is capturing the differential increase in media reporting in other countries, not the countries where the natural disaster is happening. 

For each event $E_{k(j)t}$, we define a 7-day window around the disaster’s start and end dates. This accommodates events of varying durations and accounts for both lead and lag effects. For example, if a disaster begins in country $j$ on day $t$ and ends $n$ days later, the dataset for this event, $y_{s(i)jt_k}$, includes all dates from $t_k - 7$ to $t_k + n + 7$. We then estimate the following model:

\begin{equation}
    y_{s(i)jt} = \alpha_{s(i)j} + \eta_t + \beta_{k(j),i} \times \mathbbm{1}(d = j) \times  \mathbbm{1}(t \in {t_k - \tau_l}, t_k + \tau) + \gamma_{i} + \delta_{s(i)} + \epsilon_{s(i)jt}  \label{eq:mainspec}
\end{equation}

This event study design employs a two-way fixed effects estimation that “centers” the data, removing idiosyncratic differences in reporting levels specific to each source-destination pair, around each event, while also controlling for common time shocks. More concretely, the fixed effects $\alpha_{s(i)j}$ capture baseline differences in reporting patterns from a news source $s(i)$ to a particular destination country $j$. This is important because some media outlets in country $i$ may consistently report more on country $j$ regardless of any disaster event, potentially due to linguistic familiarity, cultural ties, or historical interest. By including $\alpha_{s(i)j}$, we eliminate these persistent differences in baseline reporting, ensuring that any identified increase in coverage can be attributed to the occurrence of a natural disaster rather than long-standing reporting biases. Additionally, this centering also helps correct for possible misattributions in geographic references, which can arise due to place name ambiguities or incomplete disambiguation.

The day-level fixed effects $\eta_t$ capture common shocks or global news cycles that might influence reporting uniformly across all sources and destinations. For example, a major geopolitical event could affect worldwide news patterns on a given day. By including $\eta_t$, we control for these time-specific fluctuations, ensuring that our measured effects are not driven by day-to-day variance in global news focus.

The coefficient of interest, $\beta_{k(j),i}$, is identified through an interaction that isolates when, where, and for how long a natural disaster influences foreign reporting. Specifically, $\mathbbm{1}(j=d)$ flags the country $j$ as the one experiencing the disaster, and $\mathbbm{1}(t \in [t_k, t_k+\tau])$ indicates whether the current day $t$ falls within $\tau$ days of the disaster’s start date $t_k$. Multiplying these indicators ensures that $\beta_{k(j),i}$ measures the average change in reporting on country $j$ by media from country $i$ strictly in the wake of that specific disaster $k$. For our main analysis, we focus on $\tau=3$ days, but we also consider alternative windows, such as $\tau=7$ or $\tau=1$, to examine how quickly reporting spikes and then decays. Conceptually, $\beta_{k(j),i}$ represents the statistically detectable \emph{attention increase} that occurs within a short time frame around the disaster event, relative to a baseline period (typically the two weeks surrounding the disaster window). This helps distinguish between normal reporting fluctuations and those specifically triggered by the occurrence of a natural disaster.

It is worth noting that we estimate a complete distribution of $\beta_{k(j),i}$ coefficients across all natural disasters $k$ and for each reporting country $i$. Given the size of our dataset—up to $\emdatcount \times \countrycount = 463,587$ potential estimates—this approach yields a rich portrayal of how different disasters and reporting countries interact. When we apply a $\log(1+y)$ transformation to the dependent variable, each estimated coefficient represents the relative increase in reporting on a particular $(i,j)$ dyad after a given disaster $k$. Additionally, we employ a shrinkage procedure, setting estimates to zero if their associated t-statistic is below 1.65. Standard errors are clustered two-way by source and destination to address potential correlation in the error structure. These refined estimates $\hat{\beta}_{k(j),i}$ are then used to explore patterns and heterogeneities in media reporting increases, examining how factors such as disaster type, fatalities, and country connectedness shape the observed reporting responses.

\paragraph{Functional form} In analyzing the dependent variable—the number of articles covering a given destination country—we consider three alternative transformations to accommodate zero-inflation and overdispersion. First, we use counts in their raw form, but the sparsity and overdispersion of the dyadic data present challenges. Second, we apply a $log(1+y)$ transformation, which moderates the impact of extreme values and handles zeros naturally by adding one before taking logs. Although adding one is arbitrary, it typically yields interpretable results. Third, we consider the inverse hyperbolic sine (IHS) transformation, which, like the log transformation, manages skewness and accommodates zeros without arbitrary shifts. In practice, we find that the choice of transformation does not qualitatively affect our main findings. For clarity and ease of interpretation, we therefore focus on the $log(1+y)$ transformation.

\paragraph{Characterizing  $\hat{\beta}_{k(j),i}$} The core output of our estimation is the set of coefficients $\hat{\beta}_{k(j),i}$, each capturing the increment in media coverage from country $i$ to country $j$ following a disaster $k$. These coefficients reflect the \emph{excess attention} that arises beyond baseline reporting patterns. Because the regression includes source-destination fixed effects and day fixed effects, the resulting $\hat{\beta}_{k(j),i}$ values can be interpreted as the part of reporting that is specifically attributable to the natural disaster shock, net of persistent reporting tendencies and global time trends. In other words, they represent statistically detectable boosts in coverage that can be linked to the occurrence of a particular disaster in a given country.

\paragraph{Relaxation of the sharp event timing assumption}
While many disasters can be viewed as having a relatively clear starting point, others may have a more gradual onset or prolonged duration. To address this, we relax the assumption of a sharp event start. Instead of focusing solely on a window defined by $\tau$ days around the start date, we consider a more flexible model:

\begin{equation}
    y_{s(i)jt} = \alpha_{s(i)j} + \eta_t + \beta_{k(j),\kappa} \times \mathbbm{1}(d=j) \times  \mathbbm{1}(t = {t_k + \kappa}) + \gamma_{i} + \delta_{s(i)} + \epsilon_{s(i)jt} \label{eq:relaxed_event_timing}
\end{equation}

where $\kappa \in [-7,7]$ indexes days relative to the reported start date of the disaster. This approach allows us to trace how reporting evolves daily before, during, and after the event rather than simply aggregating over a fixed window. In this specification, we pool the coefficients $\beta_{k(j),\kappa}$ across reporting countries to simplify inference and visualize the average time profile of media attention. For comparison, we scale these flexible-time estimates so that they are directly comparable to the disaster-type-specific averages presented in Figure \ref{fig:averagereportingincrease}, allowing us to see whether the identified patterns are robust to the timing assumptions.

\subsection{Studying structure in the reporting increase estimates}
Once we have obtained the full set of $\hat{\beta}_{k(j),i}$ estimates, we conduct a series of analyses to understand their underlying structure. We explore three main dimensions of heterogeneity: (1) differences across types of natural disasters, (2) variations linked to the severity of the disaster, particularly the number of fatalities, and (3) the role of country connectedness—encompassing geographic proximity, historical ties, cultural similarities, social networks, genetic relatedness, and economic relationships—in shaping the magnitude and distribution of reporting increases. Through these exercises, we aim to identify regularities in how foreign media respond to disasters and the conditions under which attention is most pronounced.

\subsubsection{Disaster characteristics}
In examining how disaster characteristics affect reporting patterns, we focus on two key attributes: the type of disaster and the number of associated fatalities.

\paragraph{Disaster type} To analyze how different disaster types influence reporting, we first compute the average estimated increase in reporting, $\hat{\beta}_{k(j),i}$, for each disaster type. Panel A of Figure \ref{fig:averagereportingincrease} displays these averages by disaster type. In Panel B  we present reporting-country-specific results, comparing each disaster type’s average increase in reporting to that observed for earthquakes. We then fit a regression without an intercept to these relative increases. This approach demonstrates that the skew toward reporting on certain disaster types, notably earthquakes, is a broad, system-wide feature of the media landscapes represented in our data.

\paragraph{Number of fatalities} 
We also explore the extent to which the number of fatalities that are associated with an event may affect the estimated reporting increase. To do so, we compute a binned scatter plot dividing the empirical distribution of casualties across the $\emdatcount$ natural disasters into 50 bins and compute the average of the estimated reporting increase $\hat{\beta}_{k,i}$ for the natural disasters that fall into this bin. This is presented in Figure \ref{fig:deathgradient} Panel A. The pattern is robust to controlling for natural disaster type. 

To further explore differences in how countries’ media respond to disaster fatalities, we estimate the following regression model:

\begin{equation}
    \hat{\beta}_{k,i} =  \xi_i  + \nu_i \times log(\text{deaths}_k)  + \epsilon_{k,i}
\end{equation}

where $\nu_i$ represents the country-specific “death gradient” in reporting. We also estimate a version of the model without the fatalities measure, isolating $\xi_i$, which captures the average reporting increase for each reporting country, irrespective of disaster characteristics. Interpreting $\xi_i$ as a baseline scaling factor, we can relate the death gradient $\nu_i$ to this baseline. Figure \ref{fig:deathgradient} Panel B plots the ratio $\frac{\hat{\nu}_i}{\hat{\xi}_i}$ against $\hat{\xi}_i$, revealing how sensitive each country’s media is, in relative terms, to the number of fatalities. This analysis allows us to assess the importance of mortality in shaping reporting increases across different national media environments.

% Fix typo here
% This allows us to explore the hypothesis of the extent to which reporting on natural disasters appears to be a function of the   to which the increase in media reporting following a natural disaster is driven by the number of fatalities associated with the disaster.

\subsubsection{Degree of country-connectedness}
Next, we examine whether the increase in media reporting by country $i$, from a source $s(i)$, following a disaster $k(j)$ in country $j$ is influenced by how connected countries $i$ and $j$ are along various dimensions. We denote $w_{ij}$ as a measure of connectedness between countries $i$ and $j$. We consider a broad range of such measures, detailed below.

\paragraph{Measures of country connectedness}
To quantify $w_{ij}$, we consider a suite of measures capturing geographic proximity, historical and cultural ties, linguistic and religious similarities, social and genetic connectedness, preference and attitudinal overlaps, and economic linkages (notably trade). Many of these measures are drawn from the CEPII Gravity Database \citep{CEPIIgravity2022}, while others come from specialized data sources such as \citet{spolaore2024barriers}, \citet{Bailey2018FacebookSCI}, and \citet{Stojkoski2024Measuring}.

\paragraph{Geography} 
\textbf{Geographic connectedness} is represented by the great-circle distance between the most populous cities of $i$ and $j$, providing a quantitative measure of proximity. We also include \textbf{contiguity}, a binary indicator for whether two countries share a land border, based on political boundaries as of 2020. Geographic proximity can enhance information flows and mutual interest, potentially increasing cross-border reporting.

\paragraph{Historical and cultural connectedness}
We then consider a range of dyadic features that capture shared histories, languages, and cultural traits:

\begin{itemize}
    \item \textit{Same country} A binary measure of whether two countries have been part of the same political entity in the past (e.g., former colonies or unified states). The underlying data comes from CEPII data.
    \item \textit{Religious similarity}, based on indices from \citet{spolaore2024barriers}, which quantify overlap in the religious composition of the two countries. The underlying raw data stems from the World Religion Database (WRD). The measure accounts for hierarchical relationships between religions that have shared origins among Christianity, Islam, and Judaism and is weighted by the population. The measure intends to capture the expected overlap in religious affiliation between two randomly selected individuals from country $i$ and $j$ respectively.  
    \item \textit{Linguistic similarity}, based on indices from \citet{spolaore2024barriers}, which measures the degree to which the languages spoken in $i$ and $j$ are related (e.g., through shared official languages or linguistic roots). The underlying data stems from Ethnologue dataset which creates language family trees. The measure captures the "Normalized Tree Distance"  calculating the shortest path connecting two languages along the tree, scaled relative to the root. The score ranges from 0 to 1 and is further adjusted to weight for the population in a country that speaks each language.   
    \item \textit{Cultural similarity}, based on data from \citet{spolaore2024barriers}, that measures similarity of cultures as an index constructed from the World Values Survey (WVS) \citep{Inglehart2014WVS, Welzel2013Freedom}. The set of questions that are used for country-pair similarity calculation varies and is explained in more detail in their paper. We focus on the 2014 measure of cultural similarity related to the WVS from then. Countries appear more similar if the population of respondents in the WVS share a similar response profile.     
    \item \textit{Colony} A binary indicator of whether the two countries have ever had a colonial relationship with one another.    
\end{itemize}

\textbf{Social and genetic connectedness} reflects interpersonal and biological relationships. 

\textbf{Social Connectedness Index (SCI)} from \citet{Bailey2018FacebookSCI} quantifies the strength of interpersonal ties between two countries measued using data from Facebook friendships. Formally:
\[
\text{SCI}_{ij} = \frac{\text{FB Friends}_{ij}}{\text{Users}_i \times \text{Users}_j},
\]
where $\text{FB Friends}_{ij}$ is the total number of Facebook friendships between users in $i$ and $j$, and $\text{Users}_i$ and $\text{Users}_j$ are the total users in each country. We adjust this measure to focus on foreign ties by excluding same-country connections:
\[
\text{Foreign Connection Share}_{i,j} = \frac{\text{SCI}_{i,j}}{\sum_{k \neq i} \text{SCI}_{i,k}}.
\]

\textit{Genectic similarity} We include measures of \textit{genetic similarity} from \citet{Pemberton2013}, which capture the extent of shared ancestry between the populations of $i$ and $j$. We also consider \textit{genetic similarity relative to the USA}, which benchmarks the similarity of $i$ and $j$ to a common reference point, helping contextualize patterns of genetic relatedness.

To assess \textbf{preferences and attitudinal similarities}, we use data from the Global Preferences Survey (GPS) \citep{Falk2018GPS} and the Global Climate Change Survey (GCCS) \citep{andre2024climate}. The GPS provides country-level measures of six key economic preferences: patience, risk-taking, positive reciprocity, negative reciprocity, altruism, and trust.\footnote{Patience reflects time preference, measured through hypothetical choices between immediate and delayed rewards and self-assessment of willingness to delay gratification. Risk-taking captures willingness to take risks, assessed via hypothetical lottery choices and self-assessment. Positive reciprocity measures the tendency to respond kindly to friendly actions, based on scenarios involving returning favors. Negative reciprocity assesses the propensity to retaliate against unfair treatment, including costly punishment. Altruism is derived from hypothetical donation decisions and self-assessed willingness to give without expecting anything in return. Trust is based on self-reported belief in the benevolence of others' intentions.}  To measure similarity or dissimilarity we compute a pairwise Euclidean distances as:

\[
d_{ij} = \sqrt{\sum_{k=1}^{n} (x_{ik} - x_{jk})^2},
\]

where $x_{ik}$ and $x_{jk}$ are standardized values of preference $k$ for countries $i$ and $j$, and $n$ is the number of preferences. 

We proceed in a similar fashion to measure differences in climate change beliefs and policy preferences. The GCCS offers data on \textbf{climate change beliefs and preferences} from nearly 130,000 individuals in 125 countries. Key variables include willingness to pay to fight global warming, beliefs about others' willingness to contribute, social norms regarding climate action, and views on government responsibility.\footnote{Specifically, variables include: (1) Willingness to Pay (WTP) 1\%—a binary indicator if respondents are willing to contribute 1\% of their household income monthly to fight global warming; (2) Willingness to Pay Positive Amount—a binary indicator if respondents are willing to contribute a positive amount less than 1\% if not willing to contribute 1\%; (3) Willingness to Pay Index—a composite measure coding 1 for those willing to contribute at least 1\%, 0.5 for those willing to contribute a positive amount less than 1\%, and 0 for those not willing to contribute; (4) Belief about Others' WTP—respondents' estimates of how many out of 100 people in their country are willing to contribute at least 1\%; (5) Social Norms—a binary indicator if respondents believe people in their country should try to fight global warming; (6) Government Responsibility—a binary indicator if respondents think the national government should do more to fight global warming.} We aggregate responses to the country level and compute pairwise Euclidean distances between countries based on these variables.

\textbf{Economic connectedness} is captured through trade relationships. We compute measures of goods trade by calculating the average share of a country's exports that go to a destination and the average market share of imports into a country that come from a source. For digital services trade, we use estimates from \citet{Stojkoski2024Measuring}, who provide bilateral trade flows in digital products by combining corporate revenue data, consumption patterns, machine learning, and optimal transport methods.\footnote{\citet{Stojkoski2024Measuring} estimate bilateral trade in digital products by collecting corporate revenue data from large digital firms and combining it with country consumption patterns. They use machine learning to augment missing data and optimal transport methods to allocate revenues to consumption patterns, allowing them to estimate trade flows in digitally ordered and delivered products such as video games, digital advertising, and digital intermediation services.}

In leveraging these measures, we aim to capture different aspects of connectedness between countries $i$ and $j$. Geographic and ethnographic measures account for physical proximity and shared history or language, facilitating media coverage through ease of communication and mutual interest. Cultural and preference similarities reflect shared values and attitudes, potentially influencing empathy or interest in news from each other. Social connectedness indicates the strength of interpersonal relationships across borders, which may drive media reporting through shared social networks. Economic connectedness through trade reflects the interdependence of countries' economies, which might affect media attention due to economic interests. By including these diverse measures in our analysis, we comprehensively capture the factors that may influence the extent to which media in country $i$ reports on natural disasters occurring in country $j$.

\paragraph{Country connectedness and differential media reporting}
We conduct two sets of analyses to explore how country connectedness measures $w_{ij}$ relate to variations in media reporting increases. In the first set, we examine the \emph{sign} of the relationship between $w_{ij}$ and the average reporting increase $\hat{\beta}_{k(j),i}$, without conditioning on any specific disaster characteristics. This helps us understand whether certain forms of connectedness are generally associated with higher or lower media coverage, although it may not fully explain the observed heterogeneity due to potential interactions. For clarity and comparability, we standardize the features $w_{ij}$ by converting them into z-scores.

In the second set of analyses, we focus on the \emph{interaction} between $w_{ij}$ and the number of disaster-related deaths. This allows us to determine whether certain types of country connectedness modulate how strongly media coverage responds to disaster severity.

To study the \emph{sign} of $w_{ij}$ on reporting, we estimate:

\begin{equation}
    \hat{\beta}_{k(j),i} = \eta \times x_{ijk} + \nu \times w_{ij}+ \xi_{k,i} \label{eq:univariationconnectedness}
\end{equation}

To examine the \emph{interaction} between $w_{ij}$ and disaster fatalities, we estimate:

\begin{equation}
    \hat{\beta}_{k(j),i} = \eta \times x_{ijk}  + \gamma \times w_{ij} \times deaths_k +  \nu   \times w_{ij}+ \xi_{k,i} \label{eq:interactedconnectedness}
\end{equation}

Here, we estimate six different specifications of these models, each adding progressively more control variables. These controls are captured by $x_{ijk}$ and include:

\begin{itemize}

\item $x_{ijk} = \emptyset$  in this case the coefficient $\nu$ captures the average difference in the reporting increase $ \hat{\beta}_{k(j),i} $ that we can attribute to the feature $w_{ij}$ unconditionally.

\item $x_{ijk} = \alpha_j$ in this case we remove potential level differences in the reporting elasticity on countries $j$ that may be attributable, e.g. to the countries $j$ unique global position. In \cite{Fetzermediamultiplier2024} we noted for example that there are notable level differences in reporting for specific countries that may also imply differential elasticities $ \hat{\beta}_{k(j),i} $ due to more attention being devoted to disasters in country $j$, on average. 

\item $x_{ijk} = \gamma_i$ this removes potential level differences in the reporting increase in country $i$.  This may capture potential distorting effects that the specific media source sample that we have available for country $i$ may have on the reporting elasticity.

\item $x_{ijk} =\alpha_j +  \gamma_i$ this removes both the potential level differences in the estimated average reporting increase associated with the country of a news source $i$ or the country in which a disaster occurs $j$

\item $x_{ijk} = \delta_k$ removes a disaster-specific level shifter that may affect the increase in the reporting increase, on average, for all sources in a common fashion. This can be thought of as the most general way of controlling for disaster-level characteristics.

\item $s_{ijk} = \alpha_j +  \gamma_i + \delta_k$ combines all the previous additional control variables

\end{itemize}
    
The results of these analyses are presented in Appendix Figure \ref{fig:dyadic_estimate_univariate} for the \emph{sign} of $w_{ij}$, and in Figure \ref{fig:dyadic_estimate_death_interactions} for the interaction with disaster deaths. In these figures, we plot the main coefficient $\hat{\nu}$ from equation \ref{eq:univariationconnectedness} and focus on the interaction term $\hat{\gamma}$ from specification \ref{eq:interactedconnectedness}. The features have been turned into z-scores to ensure that the point estimates are more comparable. The stability of the point estimates across the specifications is a sign of the robustness of the interactive relationship to different variation. Note that in all cases, the effect of disaster deaths themselves is accounted for.

Each row in these figures displays six points, corresponding to the six specifications with increasingly stringent controls. By examining the consistency of point estimates across different sets of controls, we identify features of $w_{ij}$ that robustly correlate with media reporting increases even after accounting for multiple confounding factors. Emphasizing consistently positive or negative estimates helps us identify systematic relationships between country connectedness and cross-border media coverage of natural disasters.

\subsubsection{Fitting a random forest}

Finally, we fit random forests to the estimated $\beta_{k(j),i}$ values to assess how well the included features explain observed heterogeneity in reporting increases. This approach helps us determine the extent to which linear or non-linear interactions between dyadic connectedness measures and disaster characteristics can account for variation in media coverage.\footnote{Further, it is instructive to understand how much of the variation in the estimated $\hat{\beta}_{k(j),i}$ can be captured with the 15 combined features that we use to study the heterogeneity.}

We use the \texttt{ranger} package in R \citep{Wright2017Ranger}, training 1,000 trees with a minimum node size of 30 and at least 10 instances per terminal node. We apply scaled permutation importance to evaluate the contribution of each predictor. We fit the random forest to the estimated $\hat{\beta}_{k(j),i}$ and carry out the analysis using three sets of features separately. We run three models:

\begin{enumerate}
    \item \textbf{Dyadic Variables Only}: This model uses only the dyadic country-pair measures of connectedness as predictors listed above. We exclude variables that are not available for most country pairs. This affects the genetic similarity measures and the survey-based measures from the Global Preferences Survey \citep{Falk2018GPS} as well as the Global Climate Change Survey (GCCS) \citep{Andre2023GCCS}. This leaves us with 10 variables.
    \item \textbf{Disaster Characteristics Only}: This model uses only disaster-specific characteristics as predictors. These include the disaster type (categorical variable), the logarithm of the number of deaths, and the disaster duration in days.
    \item \textbf{Combined Model}: This model includes both the dyadic country-pair measures and the disaster characteristics as predictors, totaling 15 variables.
\end{enumerate}

\begin{landscape}
        \begin{figure}[H]
        \centering
        \caption{Measures of country-connectedness capture systematic variation in media reporting between countries when interacted with natural disaster characteristics\label{fig:country-connectedness}}
        \vspace{0.2cm}
        
                    \includegraphics[scale=0.66]{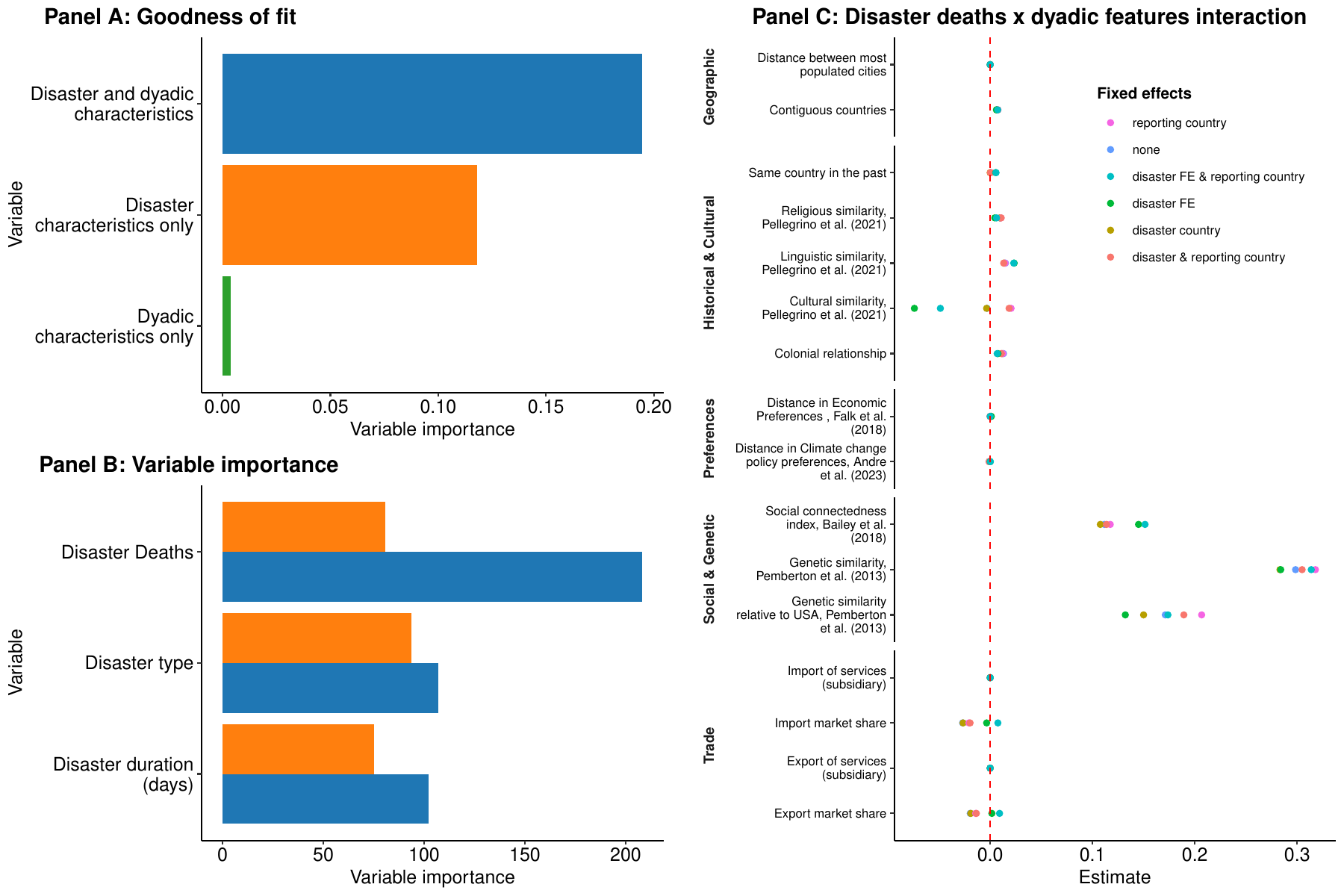}
        \captionsetup{singlelinecheck=off,font=scriptsize}
        \caption*{\textbf{Note}: Figure displays analysis of the estimated increase in dyadic media coverage across \emdatcount natural disasters that occurred since 2016. The left panel presents the results from training a random forest that aims to decompose the variation in the estimated increase in media reporting $\hat{\beta}_{k(j),i}$ of a natural disaster $k$ occurring in country $j$ through media sources associated with country $i$.  To explain the variation in this estimated reporting increase we consider either random forest that is grown using just measures of country-level connectedness labeled \emph{dyadic characteristics only}, or, we consider three natural disaster characteristics (its duration, its type and the number of deaths) associated. The third set combines both of these. The left panel presents the goodness of fit of the random forests that are grown with this set of features, while the right panel displays the variable importance of each measure. The variable importance is scaled to the maximal permutation variable importance that is measured within each model. Appendix Figure \ref{fig:dyadic_estimate_univariate} highlights the sign of the relationship between a dyadic characteristic and the extent of media coverage.}
        
        \end{figure}
        \end{landscape}

Panel A of Figure \ref{fig:country-connectedness} plots the $R^2$ of each model. Appendix Figure \ref{fig:robustnessrandomforests-variable-transformations} provides the same plot for models where the $\beta_{k(j),i}$ have been trained learned with different transformations of the dependent variable or using the more refined natural disaster-relevant article counts. We note that the pattern remains the same throughout. 

Panel B of Figure \ref{fig:country-connectedness} plots the variable importance for the main disaster variables comparing the full model that includes the dyadic characteristics with the best model that only includes the three disaster characteristics. As indicated we use permutation importance involves randomly permuting the values of a predictor variable and measuring the decrease in model performance (increase in mean squared error) resulting from the permutation. This method is less biased towards variables with many unique values or higher cardinality, making it suitable for our analysis.

We note that the increase in goodness of fit of the model that includes the dyadic characteristics is mostly driven by a sharp increase in the variable importance of the total fatalities measure, suggesting that the dyadic characteristics strongly inform the heterogeneity in reporting between countries on natural disasters in a way that is interactively shaped by the number of fatalities. 

Panel C of \ref{fig:country-connectedness} explores which dyadic characteristic appears to most notably influence the reporting differences in the interaction with the number of disaster fatalities using the linear regression model outlined in specification \ref{eq:interactedconnectedness}. In order to allow for comparisons across the different features we converted the different dyadic measures into z-scores before interacting them with the number of fatalities associated with a given natural disaster $k$. This allows for more direct comparisons. As highlighted in the discussion, the analysis suggests strongly that it is particular social- and genetic connectedness between countries that appear to shape the natural disaster deaths reporting gradient that we detected.

\subsection{Visualization in an interactive tool}
To deepen our understanding of the estimated results, we also build interactive visualizations that illustrate several core findings. We focus on two main components.

\paragraph{Visualisation of model estimates $\beta_{ji(k)}$ }
We visualize the model observations, in particular the role of fatalities in shaping reporting, by simulating a large set of disasters common to a region with different numbers of fatalities. Using these hypothetical disasters, we construct fitted values from the random forest model, namely \(\widetilde{\beta}_{ji(k)}\), for each simulated natural disaster. We then compute the average \(\bar{\beta} = \widetilde{\beta}_{ji(k)}\) for each country-pair, grouping outcomes by bins of total deaths and absorbing the type of disaster dimension. To explore the empirical distribution, we plot these values with reference to a reporting country’s percentile statistics. Specifically, let \(P_{z}(\beta_{ij} \mid j)\) indicate the \(z\)-th percentile of the empirical distribution conditional on a reporting country \(j\).

We describe two conceptual viewpoints. First, the \emph{country-of-disaster} view centers on how reporting countries \(j\) vary in their attention to a specific disaster-affected country \(i\). The data in this view are normalized within the focal disaster-affected country \(i\), across different reporting countries \(j\). This normalization adjusts for differences in absolute attention levels; for instance, some disasters may attract more global coverage due to size, media focus, or geopolitical relevance. Formally, the normalization is:
\[
\text{Normalized}_{i} \, \widehat{\beta_{ij}} 
= \frac{2 \times \bigl(\beta_{ij} - P_{5}(\beta_{ij} \,\vert\, j)\bigr)}{P_{95}(\beta_{ij} \,\vert\, j) - P_{5}(\beta_{ij} \,\vert\, j)} - 1,
\]
where \(P_{5}\) and \(P_{95}\) are the 5th and 95th percentiles, respectively.

Second, the \emph{country-of-reporting} view shifts the focus to how a single reporting country \(j\) distributes its coverage across multiple disaster-affected countries \(i\). Here, we normalize across the different countries \(i\) that are hit by disasters, allowing us to see if the same reporting country devotes disproportionate attention to certain regions or disaster types. The normalization in this view is:
\[
\text{Normalized}_{j} \, \widehat{\beta_{ij}} 
= \frac{2 \times \bigl(\beta_{ij} - P_{5}(\beta_{ij} \,\vert\, i)\bigr)}{P_{95}(\beta_{ij} \,\vert\, i) - P_{5}(\beta_{ij} \,\vert\, i)} - 1.
\]
This two-tiered perspective allows us to examine whether biases or preferences in coverage emerge in different dimensions (reporting country vs.\ disaster-affected country). For instance, a reporting country might systematically prioritize certain regions based on geopolitical, cultural, or economic ties. Conversely, a disaster-affected country might draw universal attention from most sources because of high visibility or severity. 

We also restrict 90\% of the \(\widehat{\beta}_{ij}\) observations to lie within the \([-1,1]\) range by trimming at the 5th and 95th percentiles. This avoids excessive sensitivity to extreme outliers and centers the analysis on the main mass of the distribution. In practical terms, this procedure reveals how much coverage is “typical” or “excessive” for a given scenario.

\paragraph{Path of random forests}
In addition to estimating a single fully saturated random forest, we also run a series of \(2^{10}-1\) random forests that progressively add more dyadic features alongside the baseline disaster characteristics. This best-subset selection approach traces out the path of potential random forest models, each with a distinct combination of connectedness variables. Ultimately, each specification produces an \(R^2\) measure reflecting how much additional variation in \(\hat{\beta}_{k(j),i}\) the included predictors can explain. Appendix Figure~\ref{fig:variable-random-forest-bss} shows the full distribution of these goodness-of-fit values, highlighting the role of specific features such as social connectedness, which often considerably boosts the model’s predictive accuracy.

\paragraph{Counterfactual estimates}
Finally, we aim to make the analysis more tangible by conducting \emph{counterfactual} experiments that gauge how “close” or “distant” certain disasters may appear to foreign media. Our approach builds on the observation that social and family ties between countries can alter perceptions of disaster severity. Suppose a particular country \(i\) experiences a natural disaster of a specific type with a given number of fatalities. From the perspective of a different country \(j\), one might ask how many fatalities a similar disaster in yet another country would require to elicit the same relative increase in media coverage. This question helps evaluate the relative proximity or connectedness between different countries’ populations.

\begin{figure}[H]
                \centering
                \caption{Counterfactual simulation of estimated reporting increases for hypothetical natural disasters for a storm from the perspective of German and Indian news in dataset  \label{fig:counterfactual-illustration}}
                \vspace{-0.4cm}
                \begin{subfigure}[t]{1\textwidth}
                    \centering
                    \includegraphics[scale=0.75]{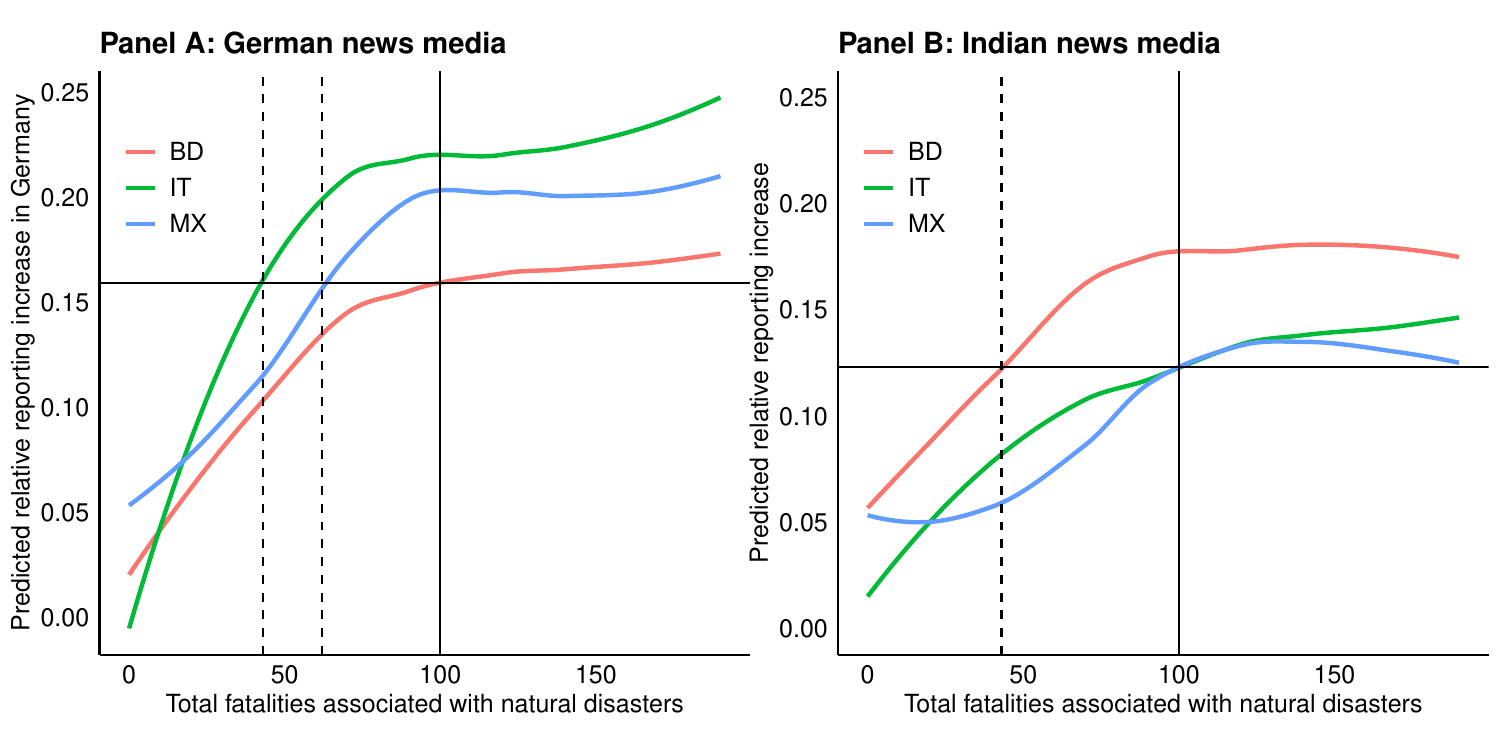}
                 %   \caption{Average reporting increase within 3 days of start of a natural disaster event}
                  
                \end{subfigure}
                \captionsetup{singlelinecheck=off,font=scriptsize}
                \caption*{\textbf{Note}: Figure presents results from  regression analysis that explores the extent to which the event-study estimated reporting increases by media associated with country $i$ reporting on country $j$ affected by a natural disaster $k$, $ \hat{\beta}_{k(j),i}$, is systematically related with various measures of country-level connectedness. In total we estimate six different specifications. We consider a feature to be a robust explanatory factor if it, across the specifications it has a consistent positive or negative sign. 
                
                }
                \label{fig:dyadic_estimate_death_interactions}
                \end{figure}

To implement this exercise, we leverage the random forest model that incorporates both disaster attributes and dyadic features. We first simulate a set of common disaster types and country locations. Next, we vary the death toll \(c\) (e.g., from 10 to 300) for each hypothetical event and compute the fitted values \(\widehat{\beta}_{ij}(c)\). This allows us to obtain a matrix of predicted coverage levels across countries and severity levels, enabling us to systematically evaluate patterns of reporting. Figure~\ref{fig:counterfactual-illustration} illustrates this concept by plotting the predicted reporting increases for storms in Bangladesh, Mexico, and Italy, respectively.

The results highlight the notion of \emph{equivalent attention}. For example, we ask: how many fatalities in one country would generate the same level of media attention as a disaster with 100 fatalities in another country? From the perspective of German media, a storm with 100 fatalities in Bangladesh elicits a similar reporting increase to a storm with approximately 43 fatalities in Italy or 62 in Mexico. Conversely, for Indian media, 45 fatalities in Bangladesh yield a level of attention comparable to that generated by 100 fatalities in either Mexico or Italy. These counterfactual scenarios make salient the often skewed nature of reporting, where equivalent attention depends not only on the raw severity of the disaster but also on the social, cultural, and geographical ties between the countries involved.

By systematically exploring these scenarios, we show that coverage patterns are highly interactive. They reflect both the absolute severity of the disaster and the relational connectedness between reporting and affected countries. Equivalent attention thus underscores how shared histories, cultural affinities, or even economic ties can modulate media responses to ostensibly similar events, highlighting non-linear effects and biases inherent in cross-border reporting.

\bibliographystyle{agsm}
\bibliography{bib.bib}

@incollection{Schafer2015Climate,
            year = {2015},
       publisher = {International Encyclopedia of the Social \& Behavioral Sciences},
          editor = {D James},
         address = {Oxford},
           title = {Climate Change and the Media},
       booktitle = {Climate Change and the Media},
           pages = {853--859},
          author = {Mike S Sch{\"a}fer},
             doi = {10.1016/B978-0-08-097086-8.91079-1},
             url = {https://www.zora.uzh.ch/id/eprint/114940/},
        language = {english}
}

@article{rinscheid2020public,
  title={What Shapes Public Support for Climate Change Mitigation Policies? The Role of Descriptive Social Norms and Elite Cues},
  author={Rinscheid, Adrian and Pianta, Silvia and Weber, Elke U.},
  journal={Behavioural Public Policy},
  volume={5},
  number={4},
  pages={503--527},
  year={2020},
  publisher={Cambridge University Press}
}

@article{merkley2020party,
  title={Party Cues in the News: Democratic Elites, Republican Backlash, and the Dynamics of Climate Skepticism},
  author={Merkley, Eric and Stecula, Dominik A.},
  journal={British Journal of Political Science},
  volume={50},
  number={3},
  pages={985--1002},
  year={2020},
  publisher={Cambridge University Press}
}

@book{Batson2011Altruism,
  title={Altruism in Humans},
  author={Batson, C. D.},
  year={2011},
  publisher={Oxford University Press}
}

@article{boykoff2007climate,
  title={Climate change and journalistic norms: A case-study of US mass-media coverage},
  author={Boykoff, Maxwell T and Boykoff, Jules M},
  journal={Geoforum},
  volume={38},
  number={6},
  pages={1190--1204},
  year={2007},
  publisher={Elsevier}
}

@book{Boykoff2011WhoSpeaksForClimate,
  title={Who Speaks for the Climate? Making Sense of Media Reporting on Climate Change},
  author={Boykoff, M. T.},
  year={2011},
  publisher={Cambridge University Press}
}

@article{Decety2014ComplexRelation,
  title={The complex relation between morality and empathy},
  author={Decety, J. and Cowell, J. M.},
  journal={Trends in Cognitive Sciences},
  volume={18},
  number={7},
  pages={337--339},
  year={2014}
}

@book{deWaal2009AgeOfEmpathy,
  title={The Age of Empathy: Nature's Lessons for a Kinder Society},
  author={de Waal, F.},
  year={2009},
  publisher={Harmony}
}

@misc{EMDAT2021,
  title={EM-DAT: The Emergency Events Database},
  author={Guha-Sapir, D.},
  howpublished={Université catholique de Louvain (UCL) - CRED, Brussels, Belgium},
  year={2021},
  note={\url{www.emdat.be}}
}

@article{Field2012Managing,
  title={Managing the risks of extreme events and disasters to advance climate change adaptation},
  author={Field, C. B. and Barros, V. and Stocker, T. F. and Dahe, Q.},
  journal={A Special Report of Working Groups I and II of the IPCC},
  year={2012}
}

@book{Goleman1995Emotional,
  title={Emotional Intelligence},
  author={Goleman, D.},
  year={1995},
  publisher={Bantam Books}
}

@article{Harcup2001NewsValues,
  title={What is news? Galtung and Ruge revisited},
  author={Harcup, T. and O'Neill, D.},
  journal={Journalism Studies},
  volume={2},
  number={2},
  pages={261--280},
  year={2001}
}

@book{Hoffman2001Empathy,
  title={Empathy and Moral Development: Implications for Caring and Justice},
  author={Hoffman, M. L.},
  year={2001},
  publisher={Cambridge University Press}
}

@book{Hulme2009WhyWeDisagree,
  title={Why We Disagree About Climate Change: Understanding Controversy, Inaction and Opportunity},
  author={Hulme, M.},
  year={2009},
  publisher={Cambridge University Press}
}

@book{IPCC2018,
  title={Global Warming of 1.5°C. An IPCC Special Report},
  author={IPCC},
  year={2018}
}

@article{Jones2022Evaluating,
  title={Evaluating the quality of disaster data: The case of EM-DAT},
  author={Jones, B. and Andrei, S. and Doyle, M. and Mumuni, E.},
  journal={International Journal of Disaster Risk Reduction},
  volume={56},
  pages={102132},
  year={2022}
}

@article{Kearney2013CrossCulturalEmpathy,
  title={Cross-cultural empathy and communication competence in healthcare providers},
  author={Kearney, G. and others},
  journal={Journal of Communication in Healthcare},
  volume={6},
  number={1},
  pages={40--49},
  year={2013}
}

@article{Lacy2001EffectsOfContent,
  title={Effects of news story content and race of source on reader perceptions},
  author={Lacy, S. and Fico, F. and Simon, T.},
  journal={Journalism \& Mass Communication Quarterly},
  volume={78},
  number={2},
  pages={285--303},
  year={2001}
}

@book{McCombs2004SettingAgenda,
  title={Setting the Agenda: The Mass Media and Public Opinion},
  author={McCombs, M.},
  year={2004},
  publisher={Polity}
}

@article{Moser2010Communicating,
  title={Communicating climate change: history, challenges, process and future directions},
  author={Moser, S. C.},
  journal={Wiley Interdisciplinary Reviews: Climate Change},
  volume={1},
  number={1},
  pages={31--53},
  year={2010}
}

@book{Painter2013Climate,
  title={Climate Change in the Media: Reporting Risk and Uncertainty},
  author={Painter, J.},
  year={2013},
  publisher={I.B. Tauris}
}

@article{Petrova2011NewspapersPoliticalCompetition,
  title={Newspapers and parties: How advertising revenues created an independent press},
  author={Petrova, M.},
  journal={American Political Science Review},
  volume={105},
  number={4},
  pages={790--808},
  year={2011}
}

@article{Pralle2009AgendaSetting,
  title={Agenda-setting and climate change},
  author={Pralle, S.},
  journal={Environmental Politics},
  volume={18},
  number={5},
  pages={781--799},
  year={2009}
}

@article{Prat2011MediaInfluence,
  title={The political economy of mass media},
  author={Prat, A. and Strömberg, D.},
  journal={Advances in Economics and Econometrics},
  volume={2},
  pages={135--187},
  year={2011}
}

@article{Schmidt2013MediaAttention,
  title={Media attention for climate change around the world: A comparative analysis of newspaper coverage in 27 countries},
  author={Schmidt, A. and Ivanova, A. and Schäfer, M. S.},
  journal={Global Environmental Change},
  volume={23},
  number={5},
  pages={1233--1248},
  year={2013}
}

@article{Soroka2019BadNews,
  title={Bad news or mad news? Sentiment scoring of negativity, fear, and anger in news content},
  author={Soroka, S. and Young, L. and Balmas, M.},
  journal={Journal of Politics},
  volume={81},
  number={1},
  pages={236--240},
  year={2019}
}

@book{Tomasello2019BecomingHuman,
  title={Becoming Human: A Theory of Ontogeny},
  author={Tomasello, M.},
  year={2019},
  publisher={Belknap Press}
}

@article{Vosoughi2018SpreadOfMisinformation,
  title={The spread of true and false news online},
  author={Vosoughi, S. and Roy, D. and Aral, S.},
  journal={Science},
  volume={359},
  number={6380},
  pages={1146--1151},
  year={2018}
}

@article{Zhuravskaya2020PoliticalEffects,
  title={Political effects of the internet and social media},
  author={Zhuravskaya, E. and Petrova, M. and Enikolopov, R.},
  journal={Annual Review of Economics},
  volume={12},
  pages={415--438},
  year={2020}
}

@article{Allcott2017SocialMediaFakeNews,
  title={Social media and fake news in the 2016 election},
  author={Allcott, H. and Gentzkow, M.},
  journal={Journal of Economic Perspectives},
  volume={31},
  number={2},
  pages={211--236},
  year={2017}
}

@article{Below2009DisasterDatabase,
  title={Disaster category classification and peril terminology for operational purposes},
  author={Below, R. and Wirtz, A. and Guha-Sapir, D.},
  journal={Working Paper},
  year={2009}
}

@article{Fan2013Ownership,
  title={Ownership consolidation and product characteristics: A study of the US daily newspaper market},
  author={Fan, Y.},
  journal={American Economic Review},
  volume={103},
  number={5},
  pages={1598--1628},
  year={2013}
}

@article{Pemberton2013,
    author = {Pemberton, Trevor J and DeGiorgio, Michael and Rosenberg, Noah A},
    title = {Population Structure in a Comprehensive Genomic Data Set on Human Microsatellite Variation},
    journal = {G3 Genes|Genomes|Genetics},
    volume = {3},
    number = {5},
    pages = {891-907},
    year = {2013},
    month = {05},
    issn = {2160-1836},
    doi = {10.1534/g3.113.005728},
    url = {https://doi.org/10.1534/g3.113.005728},
    eprint = {https://academic.oup.com/g3journal/article-pdf/3/5/891/37112047/g3journal0891.pdf},
}

@article{andre2024climate,
  author = {Andre, Pierre and Boneva, Teodora and Chopra, Farzana and Falk, Armin},
  title = {Globally Representative Evidence on the Actual and Perceived Support for Climate Action},
  journal = {Nature Climate Change},
  year = {2024}
}

@incollection{schafer2018media,
  title={Media representations of climate change: A meta-analysis of the research field},
  author={Sch{\"a}fer, Mike S and Schlichting, Inga},
  booktitle={Media Research on Climate Change},
  pages={14--32},
  year={2018},
  publisher={Routledge}
}

@book{entman2009projections,
  title={Projections of power: Framing news, public opinion, and US foreign policy},
  author={Entman, Robert M},
  year={2009},
  publisher={University of Chicago Press}
}

@article{galtung1965structure,
  title={The structure of foreign news: The presentation of the Congo, Cuba and Cyprus crises in four Norwegian newspapers},
  author={Galtung, Johan and Ruge, Mari Holmboe},
  journal={Journal of peace research},
  volume={2},
  number={1},
  pages={64--90},
  year={1965},
  publisher={Sage Publications Sage CA: Thousand Oaks, CA}
}

@book{wahl2009handbook,
  title={The handbook of journalism studies},
  author={Wahl-Jorgensen, Karin and Hanitzsch, Thomas},
  year={2009},
  publisher={Routledge New York}
}

@book{bloom2017against,
  title={Against empathy: The case for rational compassion},
  author={Bloom, Paul},
  year={2017},
  publisher={Random House}
}

@article{brewer2007importance,
  title={The importance of being we: human nature and intergroup relations.},
  author={Brewer, Marilynn B},
  journal={American psychologist},
  volume={62},
  number={8},
  pages={728},
  year={2007},
  publisher={American Psychological Association}
}

@article{simpson2015beyond,
  title={Beyond altruism: Sociological foundations of cooperation and prosocial behavior},
  author={Simpson, Brent and Willer, Robb},
  journal={Annual Review of Sociology},
  volume={41},
  number={1},
  pages={43--63},
  year={2015},
  publisher={Annual Reviews}
}

@article{choi2007coevolution,
  title={The coevolution of parochial altruism and war},
  author={Choi, Jung-Kyoo and Bowles, Samuel},
  journal={science},
  volume={318},
  number={5850},
  pages={636--640},
  year={2007},
  publisher={American Association for the Advancement of Science}
}

@book{Dryzek2011Oxford,
  title={The Oxford handbook of climate change and society},
  author={Dryzek, John S and Norgaard, Richard B and Schlosberg, David},
  year={2011},
  publisher={Oxford University Press}
}

@article{stevenson2012discursive,
  title={The discursive democratisation of global climate governance},
  author={Stevenson, Hayley and Dryzek, John S},
  journal={Environmental Politics},
  volume={21},
  number={2},
  pages={189--210},
  year={2012},
  publisher={Taylor \& Francis}
}

@article{moser2016reflections,
  title={Reflections on climate change communication research and practice in the second decade of the 21st century: what more is there to say?},
  author={Moser, Susanne C},
  journal={Wiley Interdisciplinary Reviews: Climate Change},
  volume={7},
  number={3},
  pages={345--369},
  year={2016},
  publisher={Wiley Online Library}
}

@article{o2014climate,
  title={Climate change and visual imagery},
  author={O'Neill, Saffron J and Smith, Nicholas},
  journal={Wiley Interdisciplinary Reviews: Climate Change},
  volume={5},
  number={1},
  pages={73--87},
  year={2014},
  publisher={Wiley Online Library}
}

@article{gargfetzer2024,
  title={Political Expression of Academics on Social Media},
  author={Garg, Prashant and Fetzer, Thiemo},
  journal={Forthcoming, Nature Human Behaviour},
  year={2024},
  doi={10.21203/rs.3.rs-4480504/v1},
  url={https://doi.org/10.21203/rs.3.rs-4480504/v1}
}

@article{spolaore2024barriers,
  author       = {Bruno Pelegrino and Enrico Spolaore and Romain Wacziarg},
  title        = {Barriers to Global Capital Allocation},
  journal      = {Quarterly Journal of Economics},
  note         = {Conditionally Accepted},
  year         = {2024},
}

@Preamble{ " \newcommand{\noop}[1]{} " }

@article{Gavin2009Climate,
  title={Addressing climate change: a media perspective},
  author={Gavin, Neil T},
  journal={Environmental Politics},
  volume={18},
  number={5},
  pages={765--780},
  year={2009},
  publisher={Taylor \& Francis}
}

@book{Beck2009WorldRisk,
  title={World at risk},
  author={Beck, Ulrich},
  year={2009},
  publisher={Polity}
}

@book{Decety2010Empathy,
  title={The social neuroscience of empathy},
  author={Decety, Jean and Ickes, William},
  year={2011},
  publisher={Mit press}
}

@article{DellaVigna2007FoxNews,
  title={The Fox News effect: Media bias and voting},
  author={DellaVigna, Stefano and Kaplan, Ethan},
  journal={The Quarterly Journal of Economics},
  volume={122},
  number={3},
  pages={1187--1234},
  year={2007},
  publisher={MIT Press}
}

@inproceedings{Atkinson2009NearRealTime,
  title={Near real time information mining in multilingual news},
  author={Atkinson, Martin and Van der Goot, Erik},
  booktitle={Proceedings of the 18th international conference on World wide web},
  pages={1153--1154},
  year={2009}
}

@article{Steinberger2009JRC,
  title={An introduction to the Europe Media Monitor family of applications},
  author={Steinberger, Ralf and Pouliquen, Bruno and Van der Goot, Erik},
  journal={arXiv preprint arXiv:1309.5290},
  year={2013}
}

@techreport{Burchardi2019Migrant,
  title={Migrants, ancestors, and investments},
  author={Burchardi, Konrad B and Chaney, Thomas and Hassan, Tarek A},
  year={2016},
  institution={National Bureau of Economic Research}
}

@article{Bahar2018Migration,
  title={Migration, knowledge diffusion and the comparative advantage of nations},
  author={Bahar, Dany and Rapoport, Hillel},
  journal={The Economic Journal},
  volume={128},
  number={612},
  pages={F273--F305},
  year={2018},
  publisher={Oxford University Press Oxford, UK}
}

@article{Fetzermediamultiplier2024,
  title={How big is the media multiplier? Evidence from dyadic news data},
  author={Besley, Timothy and Fetzer, Thiemo and Mueller, Hannes},
  journal={Review of Economics and Statistics},
  pages={1--45},
  year={2024},
  publisher={MIT Press One Rogers Street, Cambridge, MA 02142-1209, USA journals-info~…}
}

@article{CEPIIgravity2022,
  title={The CEPII gravity database},
  author={Conte, Maddalena and Cotterlaz, Pierre and Mayer, Thierry and others},
  year={2022},
  publisher={CEPII}
}

@article{Boykoff2009WeSpeakForTheTrees,
  title={We speak for the trees: Media reporting on the environment},
  author={Boykoff, Maxwell T},
  journal={Annual review of Environment and Resources},
  volume={34},
  number={1},
  pages={431--457},
  year={2009},
  publisher={Annual Reviews}
}

@book{Inglehart2014WVS,
  title={World Values Survey: Round Six - Country-Pooled Datafile Version},
  author={Inglehart, Ronald and Haerpfer, Christian and Moreno, Alejandro and Welzel, Christian and Kizilova, Kseniya and Diez-Medrano, Jaime and Lagos, Mart{\'\i}n and Norris, Pippa and Ponarin, Eduard and Puranen, Bi},
  year={2014},
  publisher={JD Systems Institute},
  note={Available at \url{www.worldvaluessurvey.org/WVSDocumentationWV6.jsp}}
}

@book{Welzel2013Freedom,
  title={Freedom Rising: Human Empowerment and the Quest for Emancipation},
  author={Welzel, Christian},
  year={2013},
  publisher={Cambridge University Press}
}

@article{Bailey2018FacebookSCI,
  title={Social connectedness: Measurement, determinants, and effects},
  author={Bailey, Michael and Cao, Rachel and Kuchler, Theresa and Stroebel, Johannes and Wong, Arlene},
  journal={Journal of Economic Perspectives},
  volume={32},
  number={3},
  pages={259--280},
  year={2018}
}

@article{Stojkoski2024Measuring,
  title={Estimating digital product trade through corporate revenue data},
  author={Stojkoski, Viktor and Koch, Philipp and Coll, Eva and Hidalgo, C{\'e}sar A.},
  journal={Nature Communications},
  volume={15},
  pages={5262},
  year={2024}
}

@article{Falk2018GPS,
  title={Global evidence on economic preferences},
  author={Falk, Armin and Becker, Anke and Dohmen, Thomas and Enke, Benjamin and Huffman, David and Sunde, Uwe},
  journal={The quarterly journal of economics},
  volume={133},
  number={4},
  pages={1645--1692},
  year={2018},
  publisher={Oxford University Press}
}

@article{Doulton2009TenYears,
  title={Ten years to prevent catastrophe?: Discourses of climate change and international development in the UK press},
  author={Doulton, Hugh and Brown, Katrina},
  journal={Global Environmental Change},
  volume={19},
  number={2},
  pages={191--202},
  year={2009},
  publisher={Elsevier}
}

@article{Wright2017Ranger,
  title={ranger: A fast implementation of random forests for high dimensional data in C++ and R},
  author={Wright, Marvin N and Ziegler, Andreas},
  journal={arXiv preprint arXiv:1508.04409},
  year={2015}
}

@book{Moser2007MoreBadNews,
  title={More bad news: the risk of neglecting emotional responses to climate change information.},
  author={Moser, Susanne C},
  year={2007},
  publisher={Cambridge University Press}
}

@book{Nye2004SoftPower,
  title={Soft power: The means to success in world politics},
  author={Nye, Joseph S},
  year={2004},
  publisher={Public affairs}
}

@article{IPCC2021Climate,
  title={IPCC, 2021: Climate change 2021-the physical science basis},
  author={Legg, Stephen},
  journal={Interaction},
  volume={49},
  number={4},
  pages={44--45},
  year={2021},
  publisher={Geography Teachers Association of Victoria Melbourne, Vic.}
}

@article{Coumou2012Decade,
  title={A decade of weather extremes},
  author={Coumou, Dim and Rahmstorf, Stefan},
  journal={Nature climate change},
  volume={2},
  number={7},
  pages={491--496},
  year={2012},
  publisher={Nature Publishing Group UK London}
}

@article{stern2007economics,
  title={The economics of climate change: the Stern review},
  author={Stern, Nicholas},
  journal={HM Treasury},
  year={2007}
}

@article{Gentzkow2006MediaBias,
	title        = {Media bias and reputation},
	author       = {Gentzkow, Matthew and Shapiro, Jesse M},
	year         = 2006,
	journal      = {Journal of political Economy},
	publisher    = {The University of Chicago Press},
	volume       = 114,
	number       = 2,
	pages        = {280--316}
}

@techreport{bursztyn2020misinformation,
	title        = {Misinformation during a pandemic},
	author       = {Bursztyn, Leonardo and Rao, Aakaash and Roth, Christopher P and Yanagizawa-Drott, David H},
	year         = 2020,
	institution  = {National Bureau of Economic Research}
}

@misc{jacquet2020covid,
  author       = {Jacquet, Guillaume and Verile, Marco},
  title        = {{COVID-19 news monitoring with Medical Information System (Medisys)}},
  year         = 2020,
  publisher    = {European Commission, Joint Research Centre (JRC)},
  note         = {[Dataset]},
  url          = {http://data.europa.eu/89h/bd2f71e7-0551-4f57-8e82-fcfca8c1a462}
}

\clearpage
\section{Figures and Tables}

%%%%%%%%%%%%%%%%%%%%%%%%%%%%%%%%%%%%%%%%%%%%%%%%
\begin{landscape}
\begin{figure}[H]
\centering
\caption{Increases in media reporting following a natural disaster are strongly driven by the disaster type}
    \label{fig:fig:averagereportingincrease_3days}
    \centering
    \includegraphics[scale=0.6]{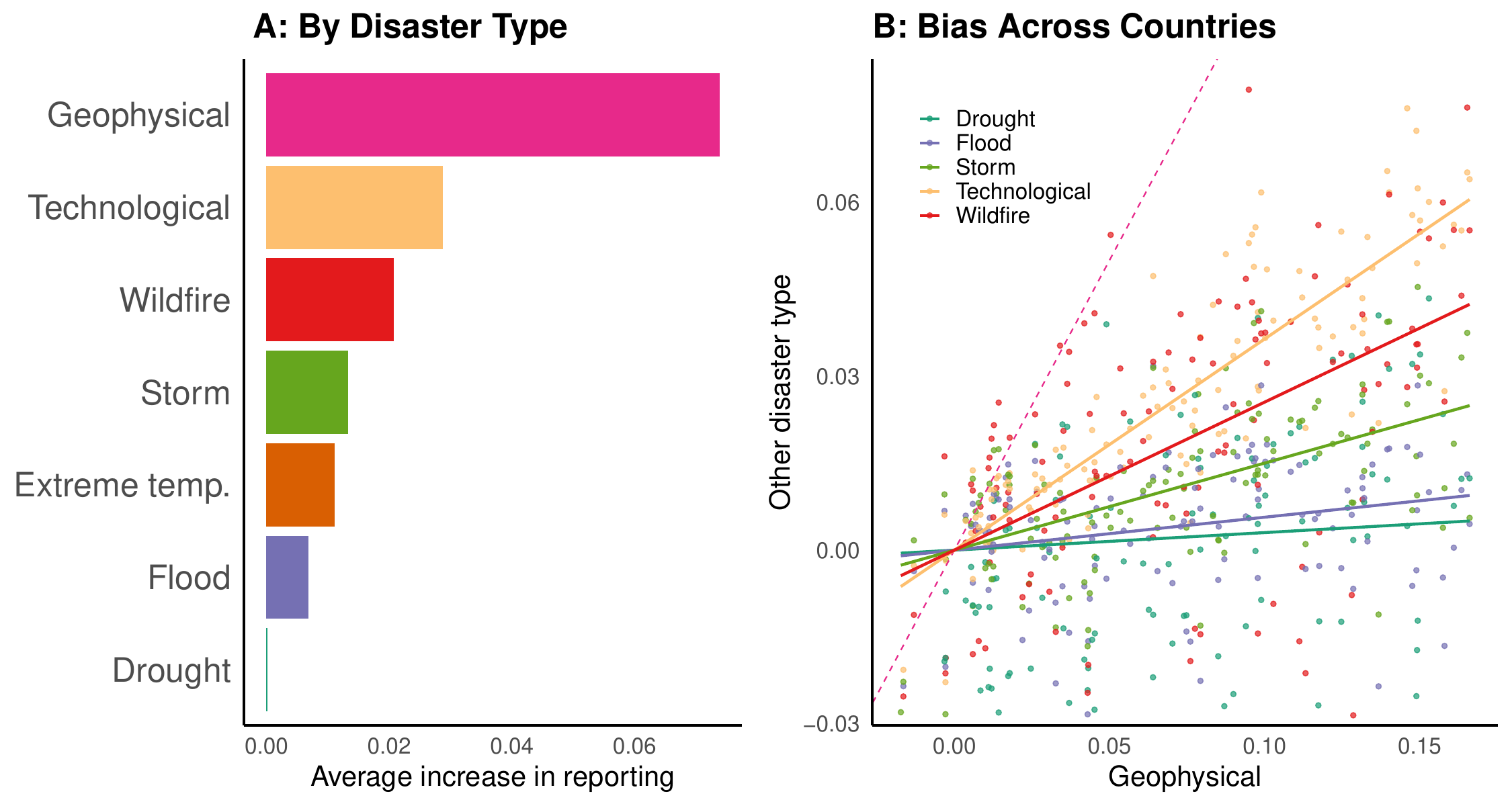}

\captionsetup{singlelinecheck=off,font=scriptsize}
\caption*{\textbf{Note}: Figure displays the estimated increase in dyadic media coverage across \emdatcount natural disasters that occurred since 2016 by natural disaster type. We construct an event-level panel dataset that captures dyadic news coverage across the \sourcecount covering \countrycount countries. For each natural disaster, ocurring in a country $i$ we construct a 7 day window around the disaster start and the disaster end date. Around this event dataset we attach dyadic daily dataset that measures the number of media articles of a news source $s$ attached to country $j(s)$ that mentions or refers to news covering any other country. Throughout we remove the distorting effects of any other time-varying shock and the average level differences in the propensity of a news source to cover any country. Panel A documents the average increase in news articles from sources $j(s)$ within 3 days after a disaster hit country $i$, relative to its news coverage of all other countries. Panel B documents the average increase in news articles from sources $j(s)$ at different points in time relative to three days before the start of the natural disaster event to allow for anticipation effects.}
\label{fig:averagereportingincrease}
\end{figure}
\end{landscape}
%%%%%%%%%%%%%%%%%%%%%%%%%%%%%%%%%%%%%%%%%%%%%%%%

%%%%%%%%%%%%%%%%%%%%%%%%%%%%%%%%%%%%%%%%%%%%%%%%
%%%%%%%%%%%%%%%%%%%%%%%%%%%%%%%%%%%%%%%%%%%%%%%%    

%%%%%%%%%%%%%%%%%%%%%%%%%%%%%%%%%%%%%%%%%%%%%%%%
%%%%%%%%%%%%%%%%%%%%%%%%%%%%%%%%%%%%%%%%%%%%%%%%        

%%%%%%%%%%%%%%%%%%%%%%%%%%%%%%%%%%%%%%%%%%%%%%%%
%%%%%%%%%%%%%%%%%%%%%%%%%%%%%%%%%%%%%%%%%%%%%%%%            

\clearpage
\appendix
\startappendixtables
\startappendixfigures

\setcounter{figure}{0}
\setcounter{table}{0}
\part*{Appendix}

\section{Inference}
% Carrying out valid inference in the context of our data-generating poses several challenges
Drawing valid inferences from our data poses several challenges. In particular, comparing patterns across the estimated effects ($\hat{\beta}_{k,i}$) is not straightforward due to (1) potential spurious correlations arising from temporal or country-specific factors, and (2) the inherent uncertainty in each estimate. This section outlines the main concerns for inference and describes the bootstrapping strategies used to address them.

\begin{itemize}
    \item \textbf{Uncertainty in the Estimated Increase in Media Reporting.} Each coefficient $\hat{\beta}_{k,i}$ measuring the increase in coverage following a disaster is subject to sampling variability, measurement error, and potential model misspecification. These sources of uncertainty can affect the reliability of the estimated effects.

    \item \textbf{Within-Reporting-Country Correlation Structures.} The propensity of news sources in country $i$ to cover foreign disasters may be systematically higher or lower for reasons unrelated to the disaster's characteristics (e.g., cultural, linguistic, or historical ties). These similarities can induce correlation in $\hat{\beta}_{k,i}$ across different $k$ and $j$, but the same reporting country $i$.

    \item \textbf{Within-Disaster-Country Correlation Structures.} Some countries experience more frequent or severe disasters due to geography or climate. Consequently, the estimates $\hat{\beta}_{k(i),j}$ linked to the same disaster-affected country $j$ may exhibit correlation unrelated to that disaster’s specific features.
\end{itemize}

To address these concerns, we evaluate the robustness of our findings to various correlation structures. To do so, we implement a bootstrapping procedure at different block levels: events, reporting countries, and disaster-affected countries. This approach allows us to gauge whether our core results hold when systematically dropping a portion of the data or focusing on specific subsets. 

\paragraph{Event-Level Bootstrapping}
We first draw 100 samples, each containing 50\% of the natural disasters recorded in EM-DAT. For each sample, we re-estimate all relevant empirical specifications, thereby obtaining a bootstrapped distribution $\{\hat{\nu}_{b,k}\}$ where $b$ indexes the bootstrap sample and $k$ indexes events. This procedure checks whether outlier or high-impact disasters could drive the main results.

\paragraph{Country-Event Block-Level Bootstrapping}
We also conduct a bootstrap in which we randomly remove (for instance) 50 disaster-affected countries $j$, thus dropping all events that occur in those countries. This accounts for the fact that the empirical distribution of disasters is not uniform across countries. Again, we replicate the main estimations on the reduced sample to see whether our findings are sensitive to the inclusion or exclusion of particular disaster-prone or high-impact countries. We then store the distribution of estimated coefficients $\{\hat{\nu}_{b,j}\}$, where $j$ indexes the removed countries.

After performing this procedure, we replicate analyses such as those used in Figure~\ref{fig:averagereportingincrease} and display the bootstrapped estimates (e.g., in Figure~\ref{fig:bootstrapped_country_averagereportingincrease_3days}). We find that the general pattern—particularly the attention gradient across different types of natural disasters—remains largely consistent, suggesting that our estimates are not driven by a small subset of countries.

\paragraph{Reporting-Country Bootstrapping}
Third, we sample among reporting countries $i$. In each bootstrap iteration, we randomly remove a subset (e.g., 50) of these reporting countries. This allows us to examine whether our results could be driven by a particular group of media ecosystems. We then re-estimate the main specifications on the reduced dataset, constructing a distribution of estimates $\{\hat{\nu}_{b,i}\}$. As with the disaster-affected country bootstrap, we find that the core findings remain stable, indicating that no single cluster of reporting countries fully explains the results.

\paragraph{Constructing Confidence Intervals and P-values}
Across each of these bootstrapping exercises, we collect the estimated coefficients and their standard errors. We take the empirical distributions of the point estimates across the approaches to construct 95\% confidence intervals around quantities such as the average increase in media reporting by disaster type. We further use this to construct p-values obtained from the empirical distributions. This procedure ensures that our inferences account for a wide range of possible dependency structures in the data.

%\section{Null effects for placebo events}                
%There may be concerns that the results could be capturing just idiosyncratic patterns. To address this concern we carry out a permutation exercise with a sample of 500 fake-events that are generated in a way that matches empirical moments of the underlying natural disasters dataset. Specifically, we generate random disaster types and random dates within each year, and then estimates the main specification \ref{eq:mainspec} to see to what extent we observe a pattern in the post-event time window. 

%Figure \ref{fig:bootstrapped_placebo_averagereportingincrease_3days} presents the results of this exercise. We find that the patterns detected in Figure \ref{fig:averagereportingincrease} are not present in the placebo events.

\section{Additional figures and tables}

%%%%%%%%%%%%%%%%%%%%%%%%%%%%%%%%%%%%%%%%%%%%%%%%            
 \begin{figure}[H]
    \centering
    \caption{Robustness to alternative dependent variables, functional forms and time-windows}
%    \vspace{+0.4cm}

    \text{\textbf{Panel A}: Dyadic counts (level)} \\
    \begin{subfigure}[t]{0.38\textwidth}
        \caption{within 1 day of event start date}
    %   \label{fig:fig:averagereportingincrease_3days}
        \includegraphics[width=\textwidth]{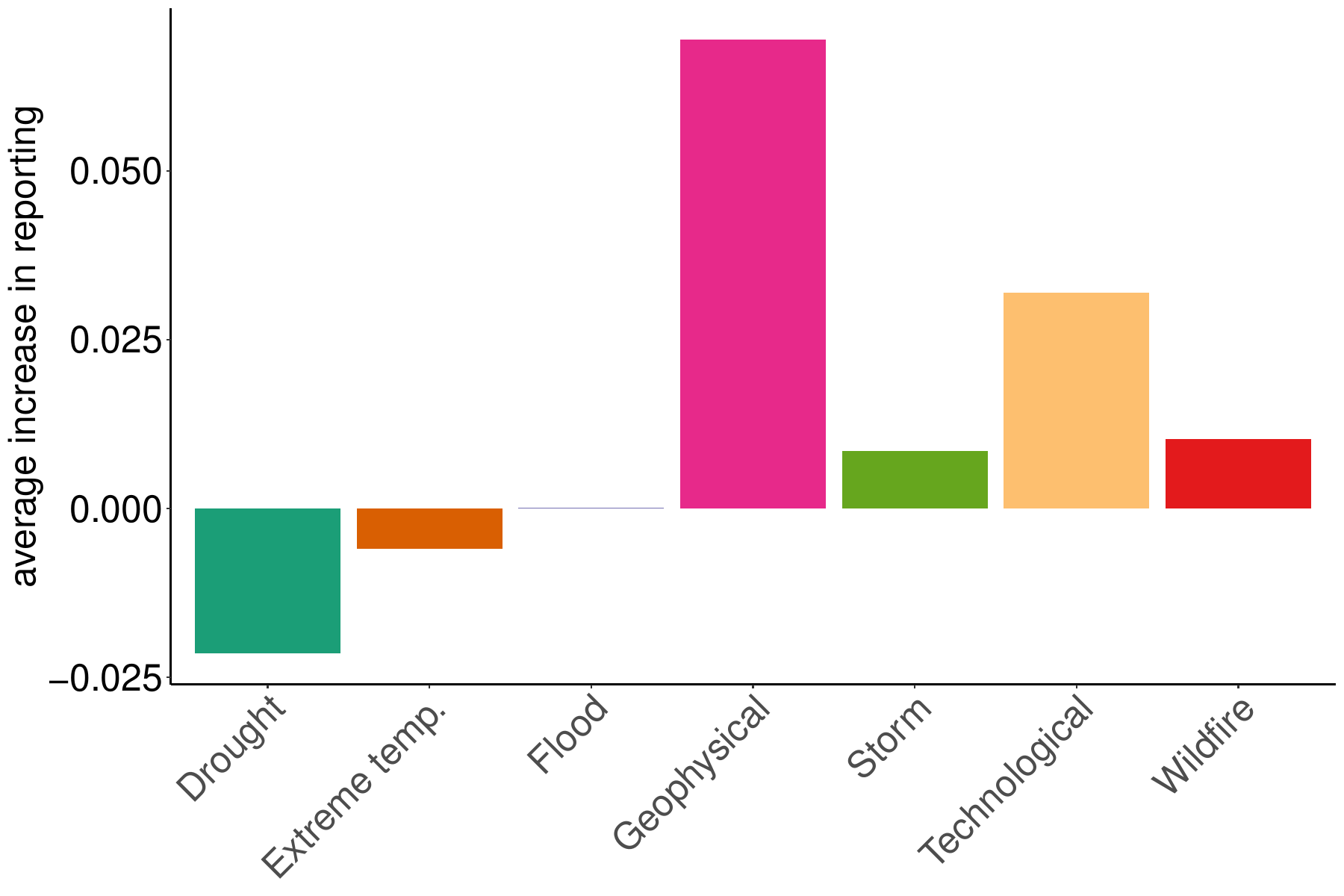}
        
    \end{subfigure}
    \hfill
    \begin{subfigure}[t]{0.38\textwidth}
        \caption{within 3 days of event start date }
    %    \label{fig:averagereportingincrease_flexible}
        \centering
        \includegraphics[width=\textwidth]{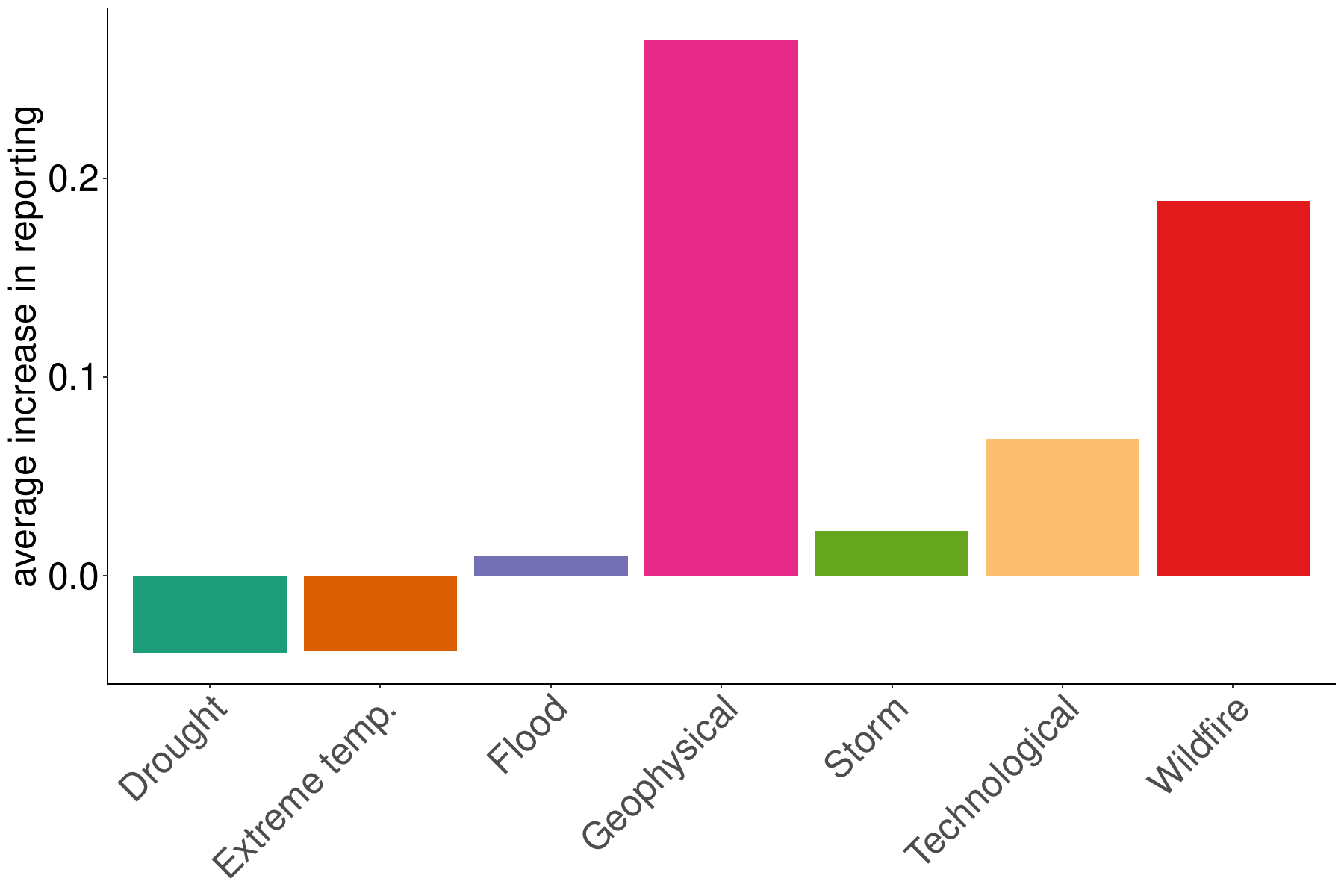}
    \end{subfigure}

     \text{\textbf{Panel B}: log(dyadic counts)} \\
    \begin{subfigure}[t]{0.38\textwidth}
        \caption{within 1 day of event start date}
   %     }\label{fig:fig:averagereportingincrease_3days}
        \includegraphics[width=\textwidth]{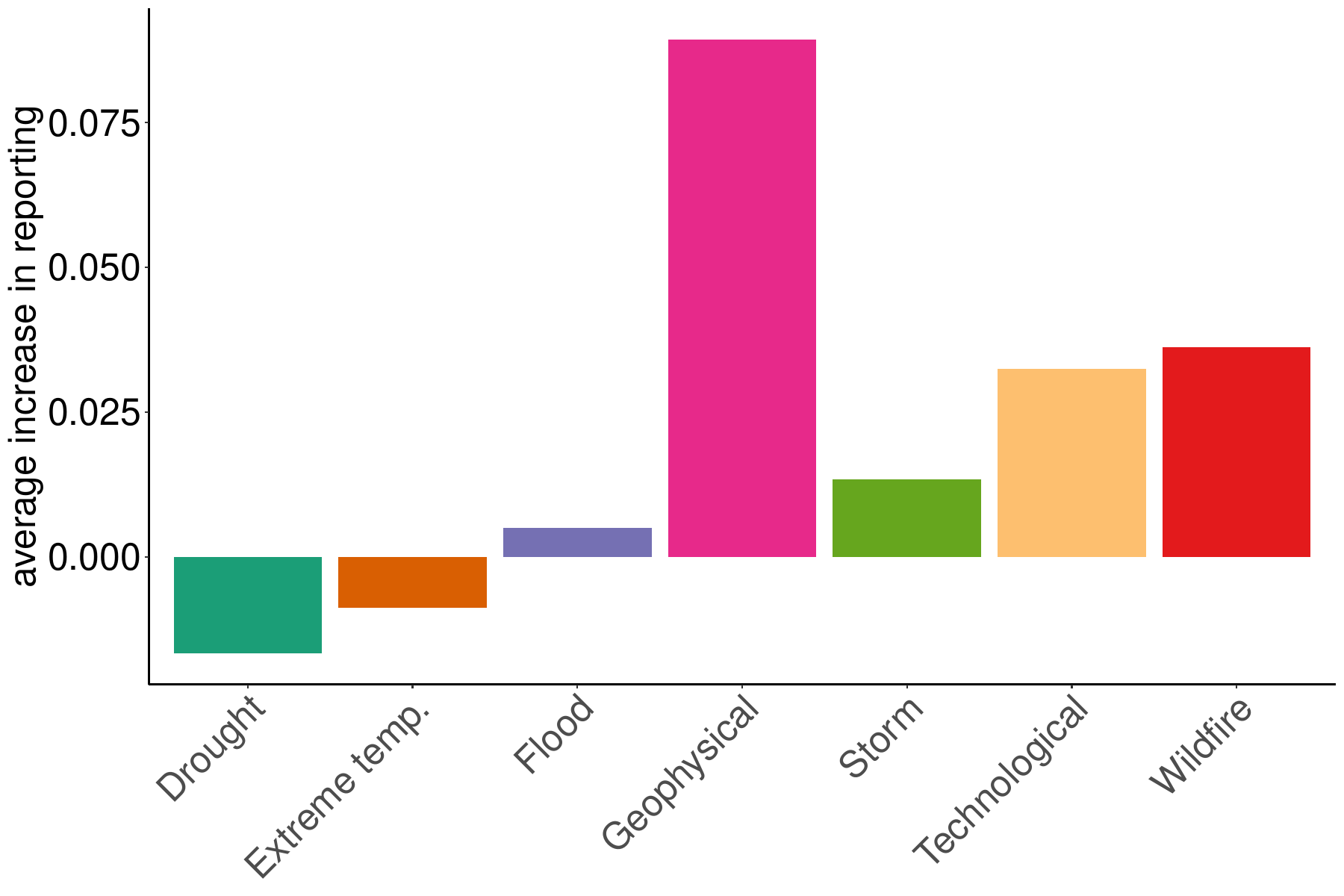}
    \end{subfigure}
    \hfill
    \begin{subfigure}[t]{0.38\textwidth}
        \caption{within 3 days of event start date }
   %     \label{fig:averagereportingincrease_flexible}
        \centering
        \includegraphics[width=\textwidth]{figures/estimate-average-event-custom_disaster_type-ihs-dyadic_counts-3.pdf}
    \end{subfigure}
 
    \text{\textbf{Panel C}: ihs(dyadic counts)} \\
    \begin{subfigure}[t]{0.38\textwidth}
        \caption{within 1 day of event start date}
   %     }\label{fig:fig:averagereportingincrease_3days}
        \includegraphics[width=\textwidth]{figures/estimate-average-event-custom_disaster_type-ihs-dyadic_counts-3.pdf}
    \end{subfigure}
    \hfill
    \begin{subfigure}[t]{0.4\textwidth}
        \caption{within 3 days of event start date }
   %     \label{fig:averagereportingincrease_flexible}
        \centering
        \includegraphics[width=\textwidth]{figures/estimate-average-event-custom_disaster_type-ihs-dyadic_counts-3.pdf}
    \end{subfigure}

    \captionsetup{singlelinecheck=off,font=scriptsize}
    \caption*{\textbf{Note}: Figure displays the estimated increase in dyadic media coverage across \emdatcount natural disasters that occurred since 2016 by natural disaster type. We construct an event-level panel dataset that captures dyadic news coverage across the \sourcecount covering \countrycount countries. For each natural disaster, ocurring in a country $i$ we construct a 7 day window around the disaster start and the disaster end date. Around this event dataset we attach dyadic daily dataset that measures the number of media articles of a news source $s$ attached to country $j(s)$ that mentions or refers to news covering any other country. Throughout we remove the distorting effects of any other time-varying shock and the average level differences in the propensity of a news source to cover any country. Across the panels different measures of media coverage are considered. Throughout this figure we focus on any dyadic counts referring to any media coverage that mentions the country name or a location within that country that was affected by a natural disaster and the country of origin on the media outlet. The left column focuses on a single day after the event start, while the right column focuses on a three day event window after event start.}
    \label{fig:averagereportingincrease-dyadic-counts}
    \end{figure}
%%%%%%%%%%%%%%%%%%%%%%%%%%%%%%%%%%%%%%%%%%%%%%%%            

%%%%%%%%%%%%%%%%%%%%%%%%%%%%%%%%%%%%%%%%%%%%%%%%            
\begin{figure}[H]
\centering
\caption{Average reporting increase with flexible event window}
\vspace{0.4cm}

    \includegraphics[width=\textwidth]{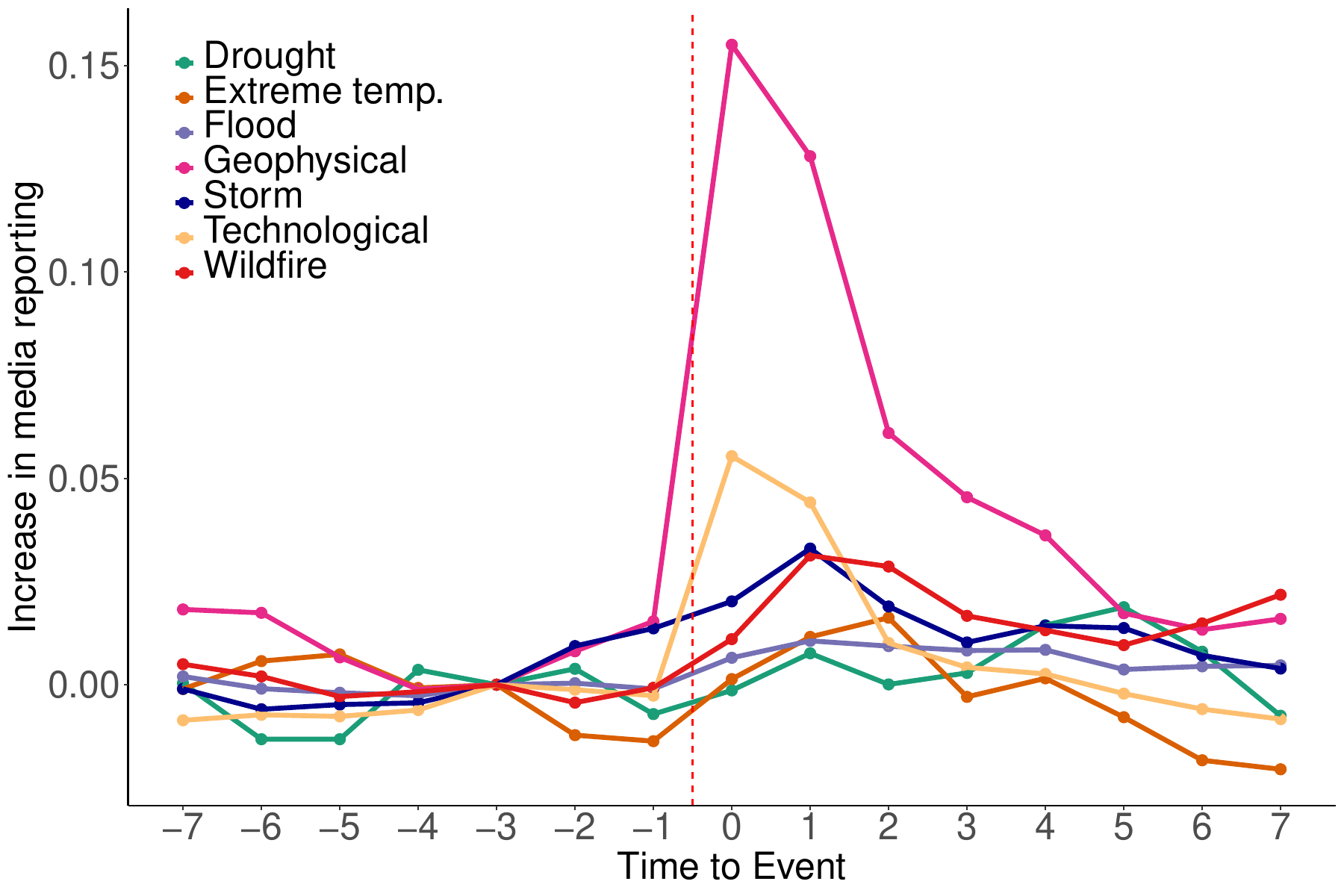}
\captionsetup{singlelinecheck=off,font=scriptsize}
\caption*{\textbf{Note}: Figure displays the estimated increase in dyadic media coverage across \emdatcount natural disasters that occurred since 2016 by natural disaster type. We construct an event-level panel dataset that captures dyadic news coverage across the \sourcecount covering \countrycount countries. For each natural disaster, ocurring in a country $i$ we construct a 7 day window around the disaster start and the disaster end date. Around this event dataset we attach dyadic daily dataset that measures the number of media articles of a news source $s$ attached to country $j(s)$ that mentions or refers to news covering any other country. Throughout we remove the distorting effects of any other time-varying shock and the average level differences in the propensity of a news source to cover any country. Panel A documents the average increase in news articles from sources $j(s)$ within 3 days after a disaster hit country $i$, relative to its news coverage of all other countries. Panel B documents the average increase in news articles from sources $j(s)$ at different points in time relative to three days before the start of the natural disaster event to allow for anticipation effects.}
\label{fig:averagereportingincrease_flexible}
\end{figure}
%%%%%%%%%%%%%%%%%%%%%%%%%%%%%%%%%%%%%%%%%%%%%%%%            

%%%%%%%%%%%%%%%%%%%%%%%%%%%%%%%%%%%%%%%%%%%%%%%%
\begin{figure}[H]
\centering
\caption{Measures of country-level connectedness impact the statistically detectable reporting decreases following a natural disaster}
\vspace{-0.4cm}
\begin{subfigure}[t]{1\textwidth}
\centering
\includegraphics[scale=0.8]{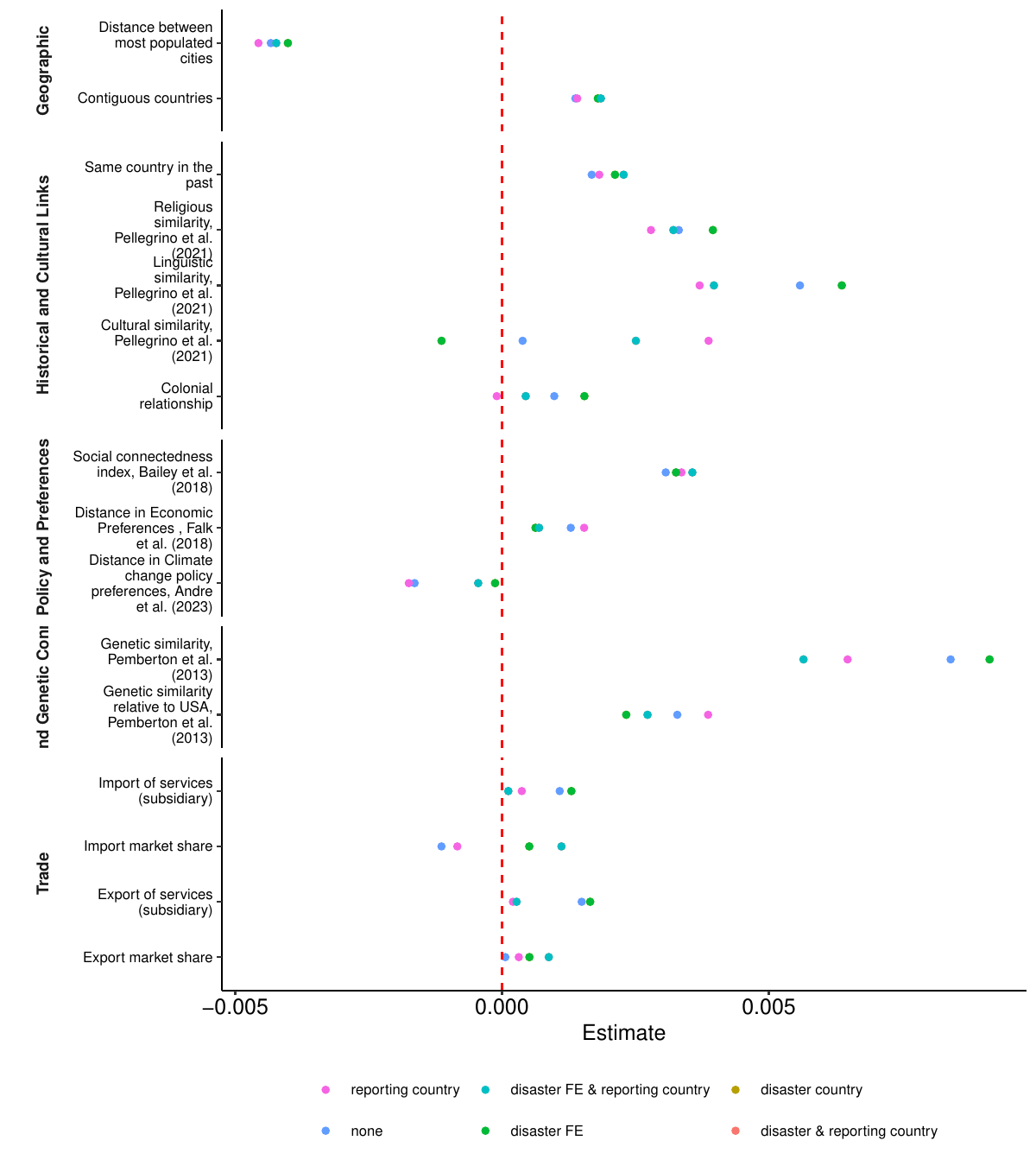}
                 %   \caption{Average reporting increase within 3 days of start of a natural disaster event}
                  %  \label{fig:bootstrapped_averagereportingincrease_3days}
\end{subfigure}
\captionsetup{singlelinecheck=off,font=scriptsize}
\caption*{\textbf{Note}: Figure presents results from  regression analysis that explores the extent to which the event-study estimated reporting increases by media associated with country $i$ reporting on country $j$ affected by a natural disaster $k$, $ \hat{\beta}_{k(j),i}$, is systematically related with various measures of country-level connectedness. In total we estimate six different specifications. We consider a feature to be a robust explanatory factor if it, across the specifications it has a consistent positive or negative sign.           
                }
\label{fig:dyadic_estimate_univariate}
\end{figure}
%%%%%%%%%%%%%%%%%%%%%%%%%%%%%%%%%%%%%%%%%%%%%%%%            

%%%%%%%%%%%%%%%%%%%%%%%%%%%%%%%%%%%%%%%%%%%%%%%%
\begin{figure}[H]
\centering
\caption{Random forest model saturation evaluated using best-subset selection approach}
\vspace{-0.4cm}
\begin{subfigure}[t]{1\textwidth}
\centering
\includegraphics[scale=0.35]{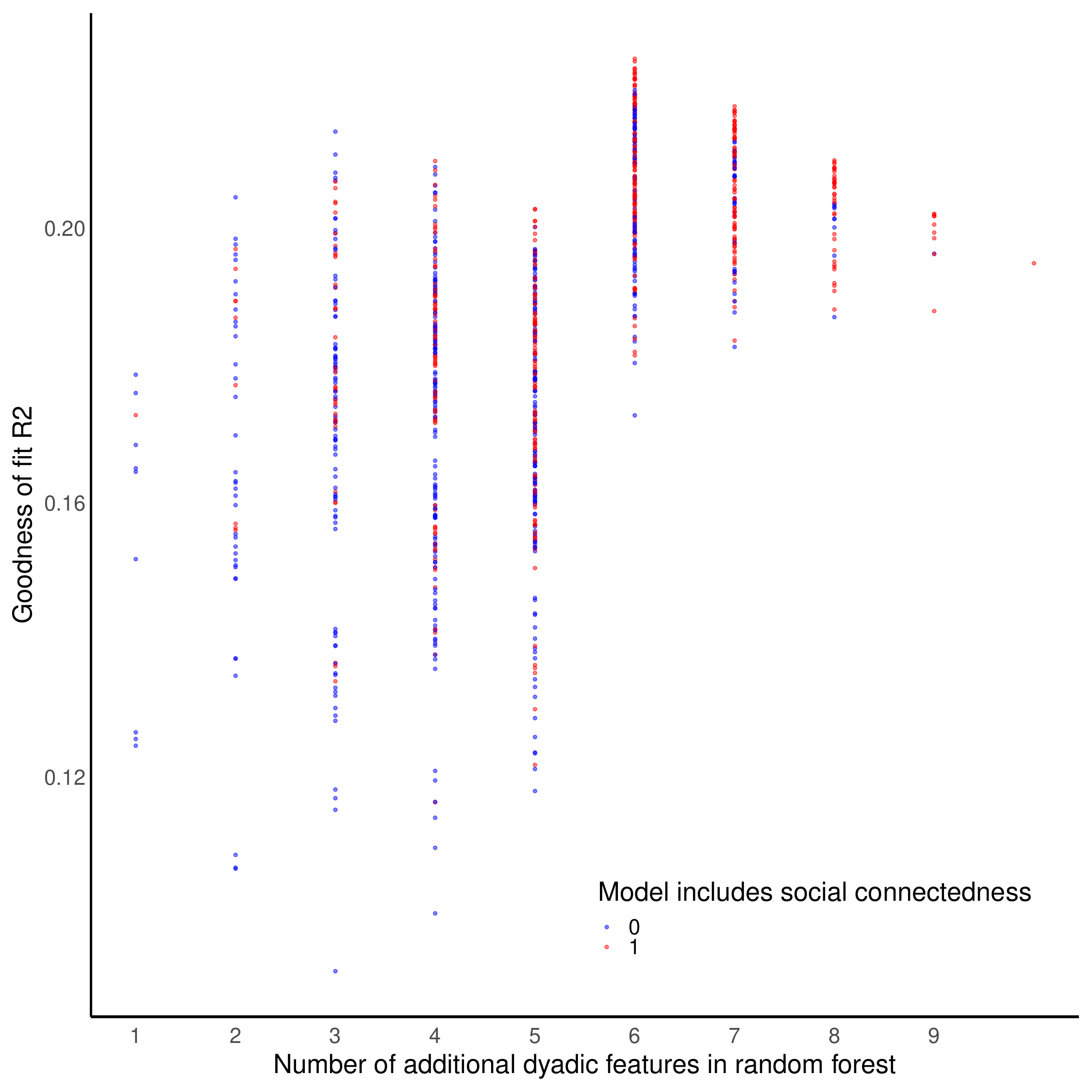}
                 %   \caption{Average reporting increase within 3 days of start of a natural disaster event}
                  %  \label{fig:variable-random-forest-bss}
\end{subfigure}
\captionsetup{singlelinecheck=off,font=scriptsize}
\caption*{\textbf{Note}: Figure presents the R2 that is obtained estimating 1023 random forests models that successively add more dyadic features. In total there are 10 additional features in addition to the pure disaster characteristics. The R2 plotted out at the x-axis value of 1 captures the R2 that is associated with the 10   presents results from  regression analysis that explores the extent to which the event-study estimated reporting increases by media associated with country $i$ reporting on country $j$ affected by a natural disaster $k$, $ \hat{\beta}_{k(j),i}$, is systematically related with various measures of country-level connectedness. In total we estimate six different specifications. We consider a feature to be a robust explanatory factor if it, across the specifications it has a consistent positive or negative sign. 
                
                }
\label{fig:dyadic_estimate_univariate}
\end{figure}
%%%%%%%%%%%%%%%%%%%%%%%%%%%%%%%%%%%%%%%%%%%%%%%%            

%%%%%%%%%%%%%%%%%%%%%%%%%%%%%%%%%%%%%%%%%%%%%%%%
\begin{landscape}
\begin{figure}[t]
    \centering
    \caption{Robustness to alternative transformations of the dependent variable\label{fig:robustnessrandomforests-variable-transformations}}
    \vspace{0.2cm}

    \renewcommand{\arraystretch}{1.5} % Adjust row height
    \setlength{\tabcolsep}{8pt} % Adjust column spacing

    \begin{tabular}{ccc}
        \multicolumn{3}{c}{\textbf{\emph{Panel A: Dependent variable is dyadic counts using the following transformations}}} \\[0.5em]
        \textbf{$\log + 1$} & IHS & Counts \\

        \includegraphics[scale=0.35]{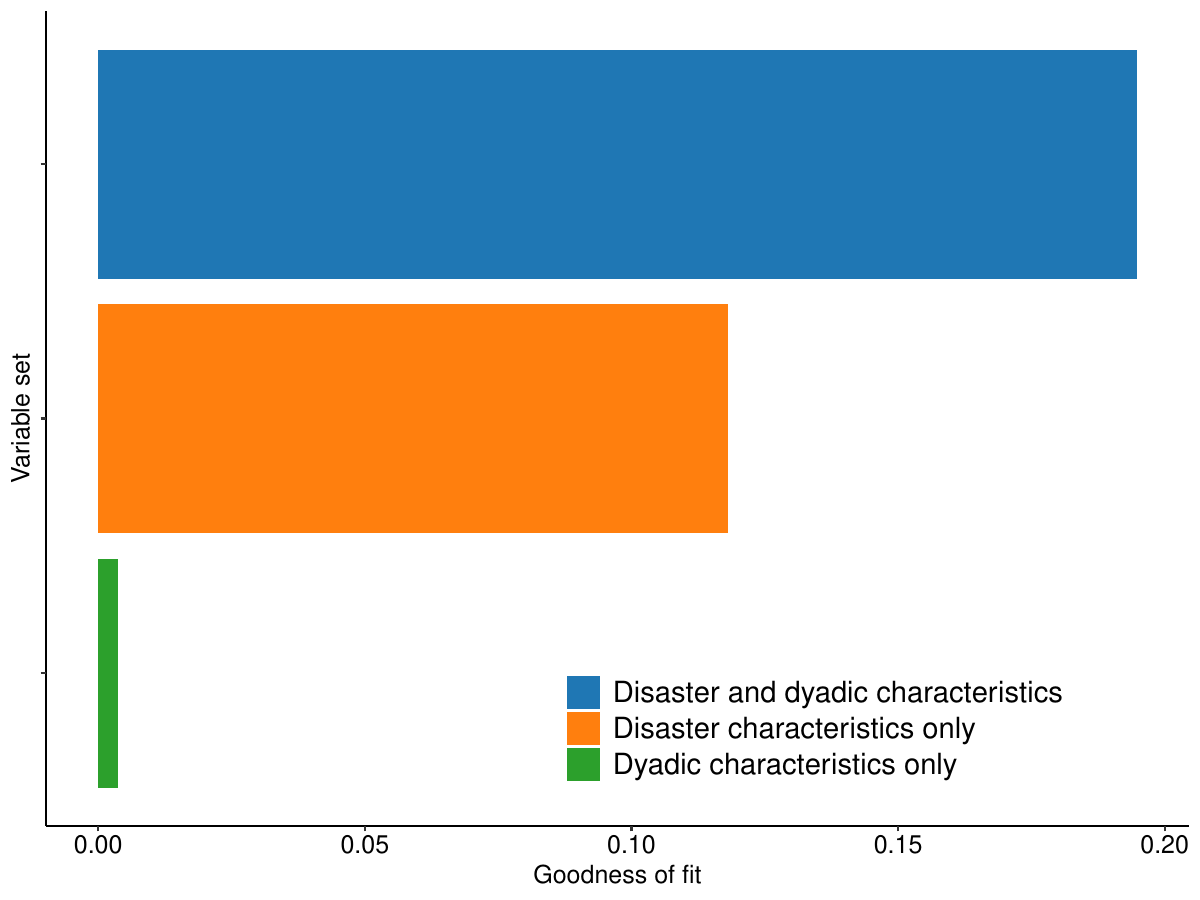} &
        \includegraphics[scale=0.35]{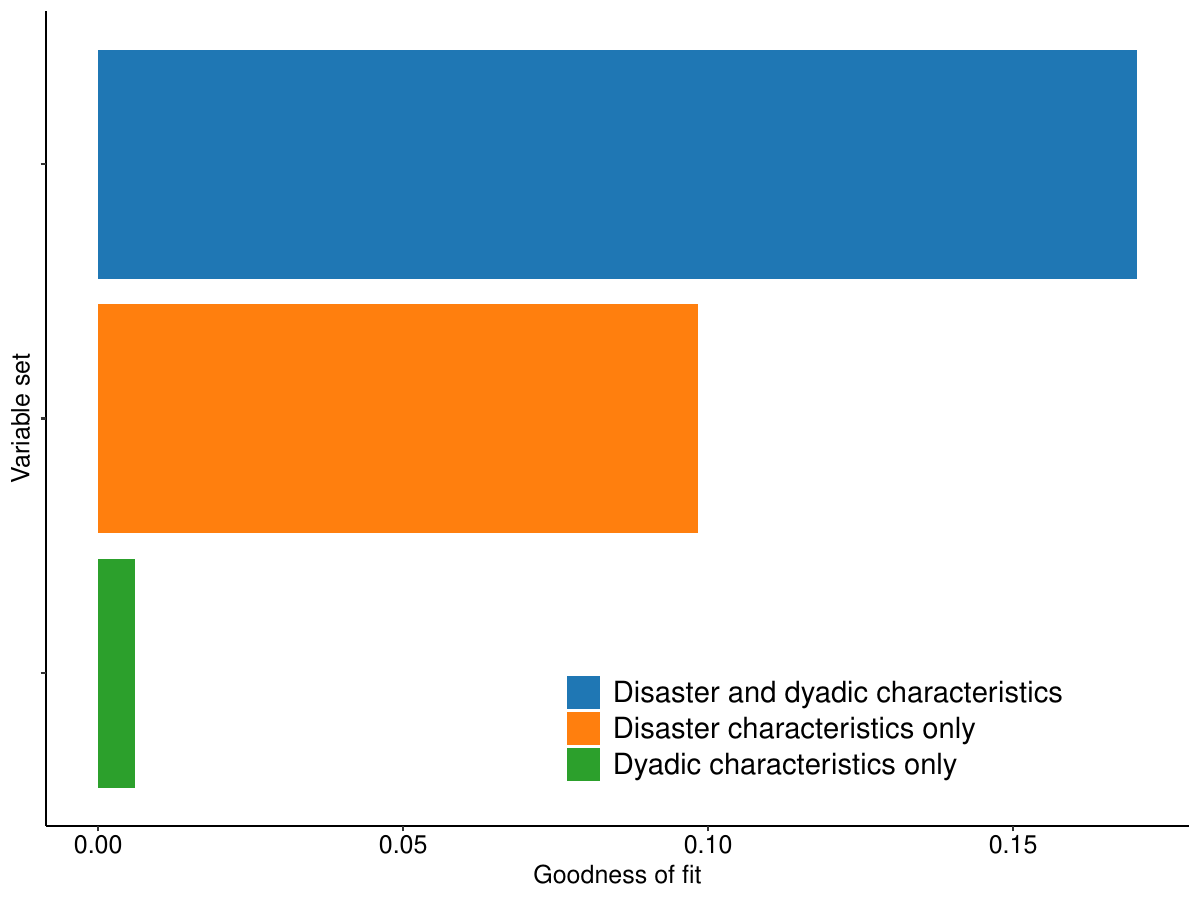} &
        \includegraphics[scale=0.35]{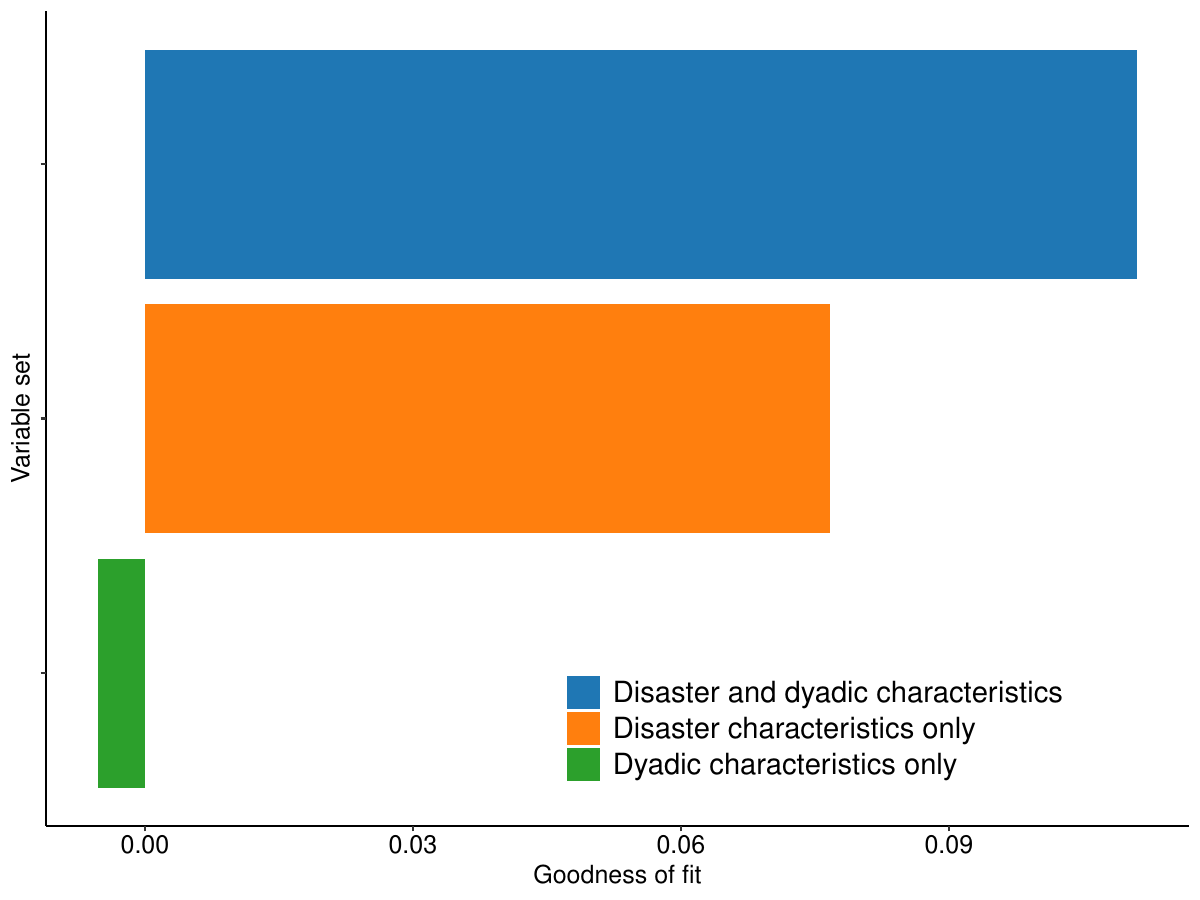} \\[1em]

        \multicolumn{3}{c}{\textbf{\emph{Panel B: Dependent variable is natural disaster mentioning counts using the following transformations}}} \\[0.5em]
        \textbf{$\log + 1$} & IHS  & Counts \\

        \includegraphics[scale=0.35]{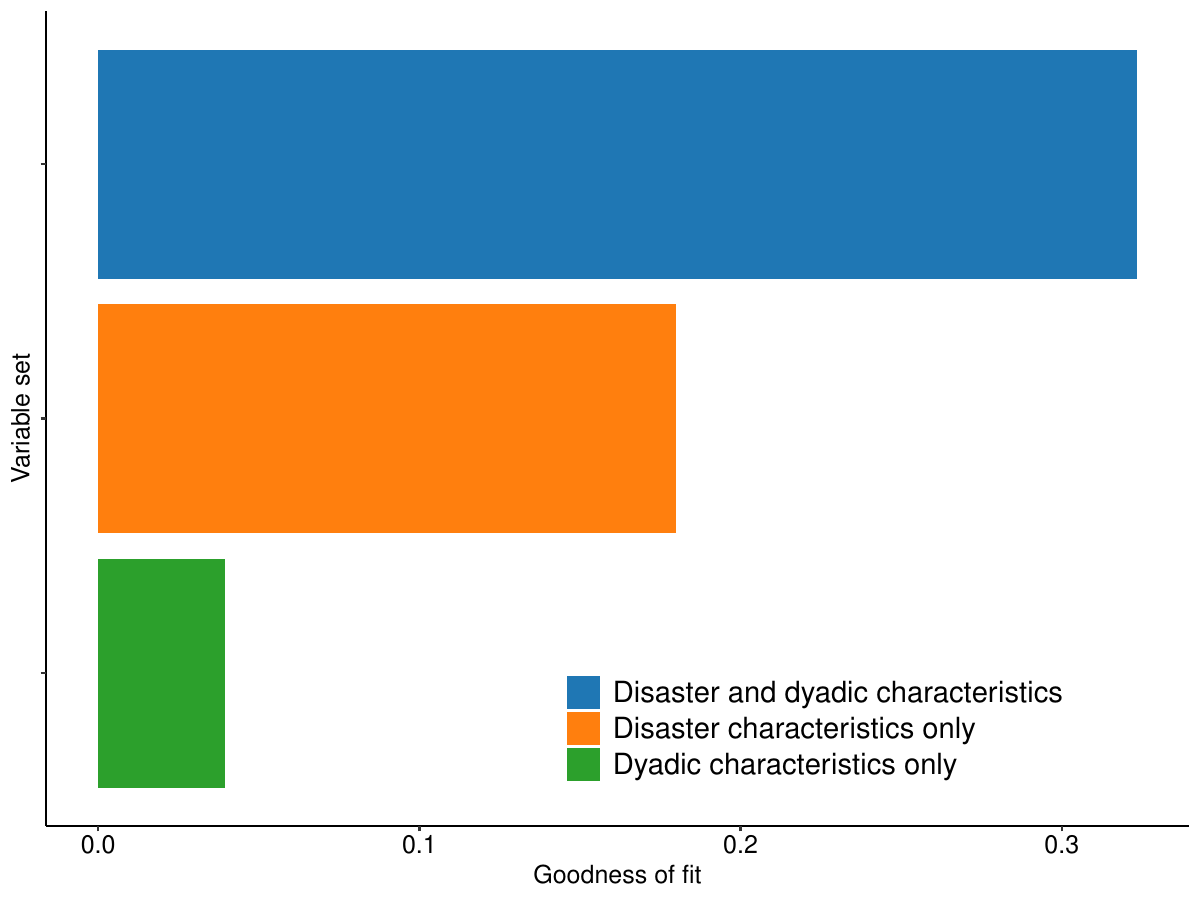} &
        \includegraphics[scale=0.35]{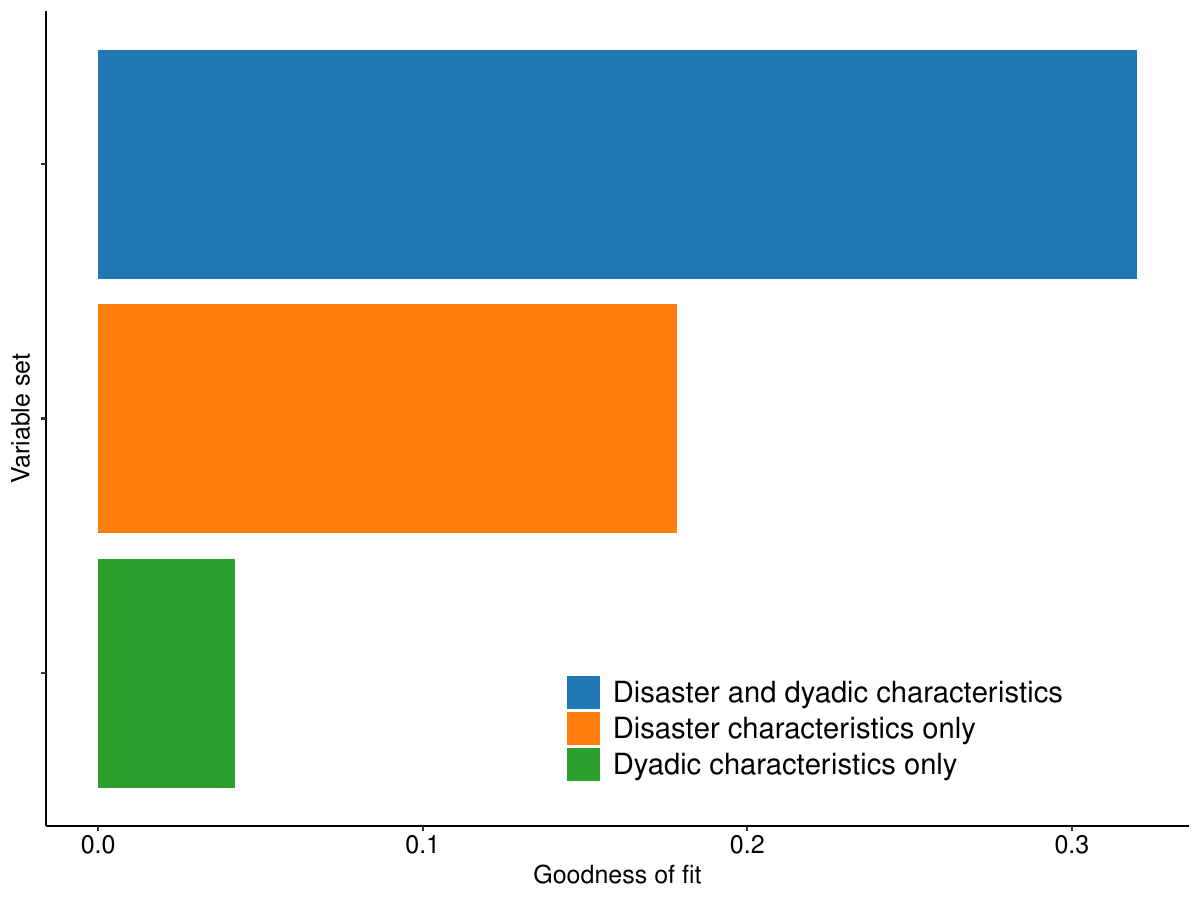} &
        \includegraphics[scale=0.35]{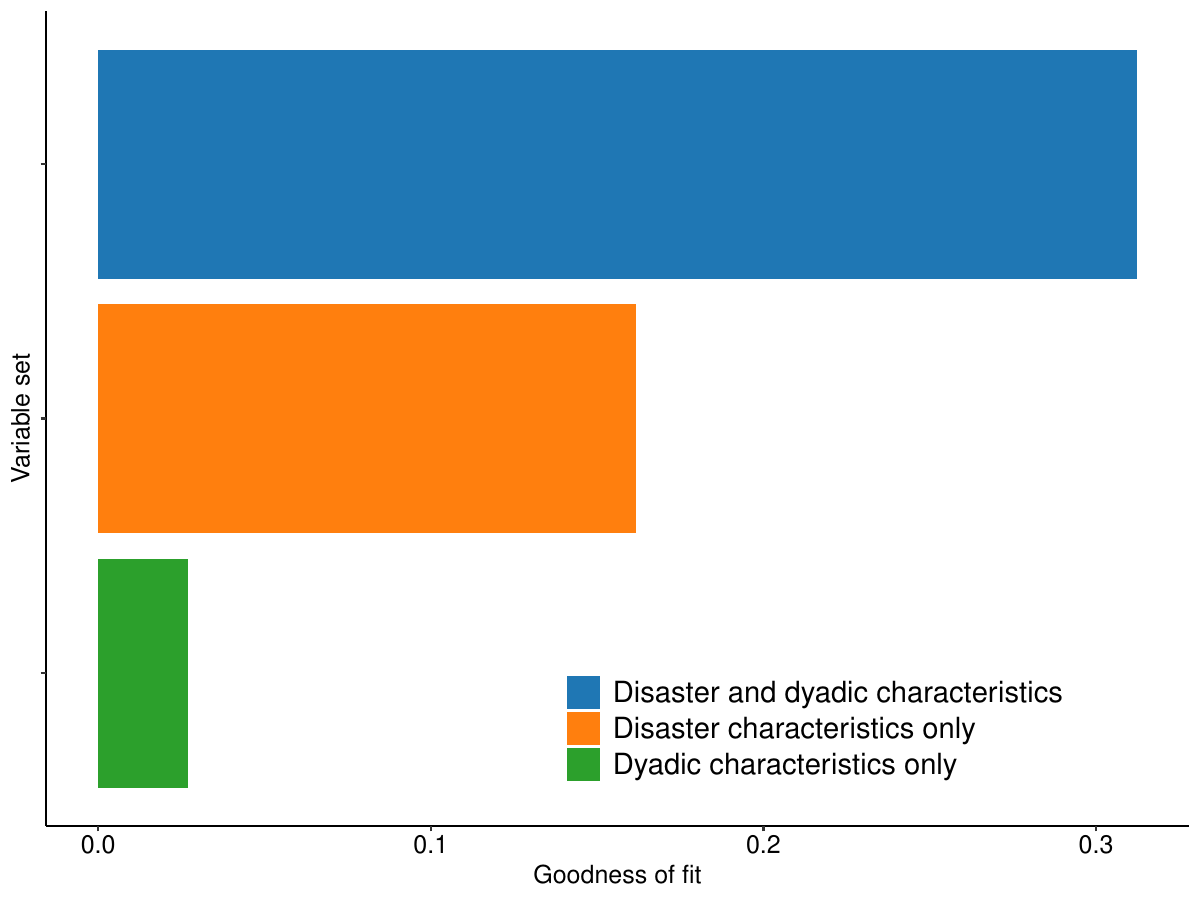} \\
    \end{tabular}

    \captionsetup{singlelinecheck=off, font=scriptsize}
    \caption*{\textbf{Note}: Figure displays analysis of the estimated increase in dyadic media coverage across \emdatcount natural disasters that occurred since 2016. The left panel presents the results from training a random forest that aims to decompose the variation in the estimated increase in media reporting $\hat{\beta}_{k(j),i}$ of a natural disaster $k$ occurring in country $j$ through media sources associated with country $i$. To explain the variation in this estimated reporting increase, we consider either random forests grown using just measures of country-level connectedness (labeled \emph{dyadic characteristics only}) or three natural disaster characteristics (its duration, type, and the number of deaths) associated with the event. The third set combines both of these. The left panel presents the goodness of fit of the random forests grown with this set of features, while the right panel displays the variable importance of each measure. The variable importance is scaled to the maximal permutation variable importance measured within each model. Appendix Figure \ref{fig:dyadic_estimate_univariate} highlights the sign of the relationship between a dyadic characteristic and the extent of media coverage.}
\end{figure}
\end{landscape}

%%%%%%%%%%%%%%%%%%%%%%%%%%%%%%%%%%%%%%%%%%%%%%%%        

\begin{figure}[H]
\centering
\caption{Number of Sources By Language }
\vspace{-0.4cm}
\centering
\includegraphics[width=\textwidth]{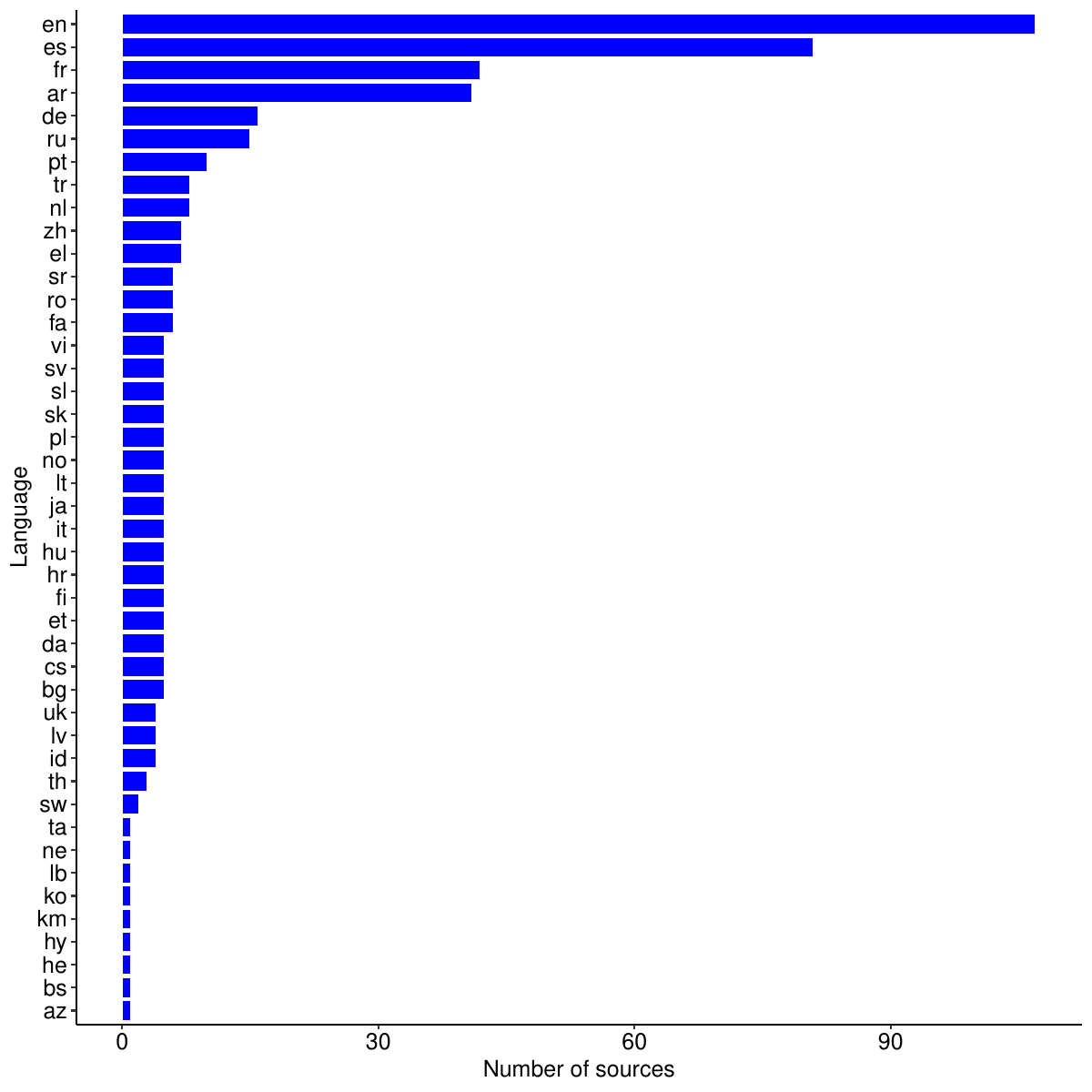}
                \label{fig:sources_by_language}
\captionsetup{singlelinecheck=off,font=scriptsize}
\caption*{\textbf{Note}: Figure tabulates the number of sources by the respective language. In total there are 44 different languages in the main dataset covering the \sourcecount sources across \countrycount countries. The EMM machine translates content to English.  }
\end{figure}
            
%%%%%%%%%%%%%%%%%%%%%%%%%%%%%%%%%%%%%%%%%%%%%%%%%%%%%%%%%%%%%%%%%%%%%
    
\begin{figure}[H]
\centering
\caption{Differences in increases in media reporting following a natural disaster by media ownership}
\vspace{-0.4cm}
\centering
\includegraphics[width=\textwidth]{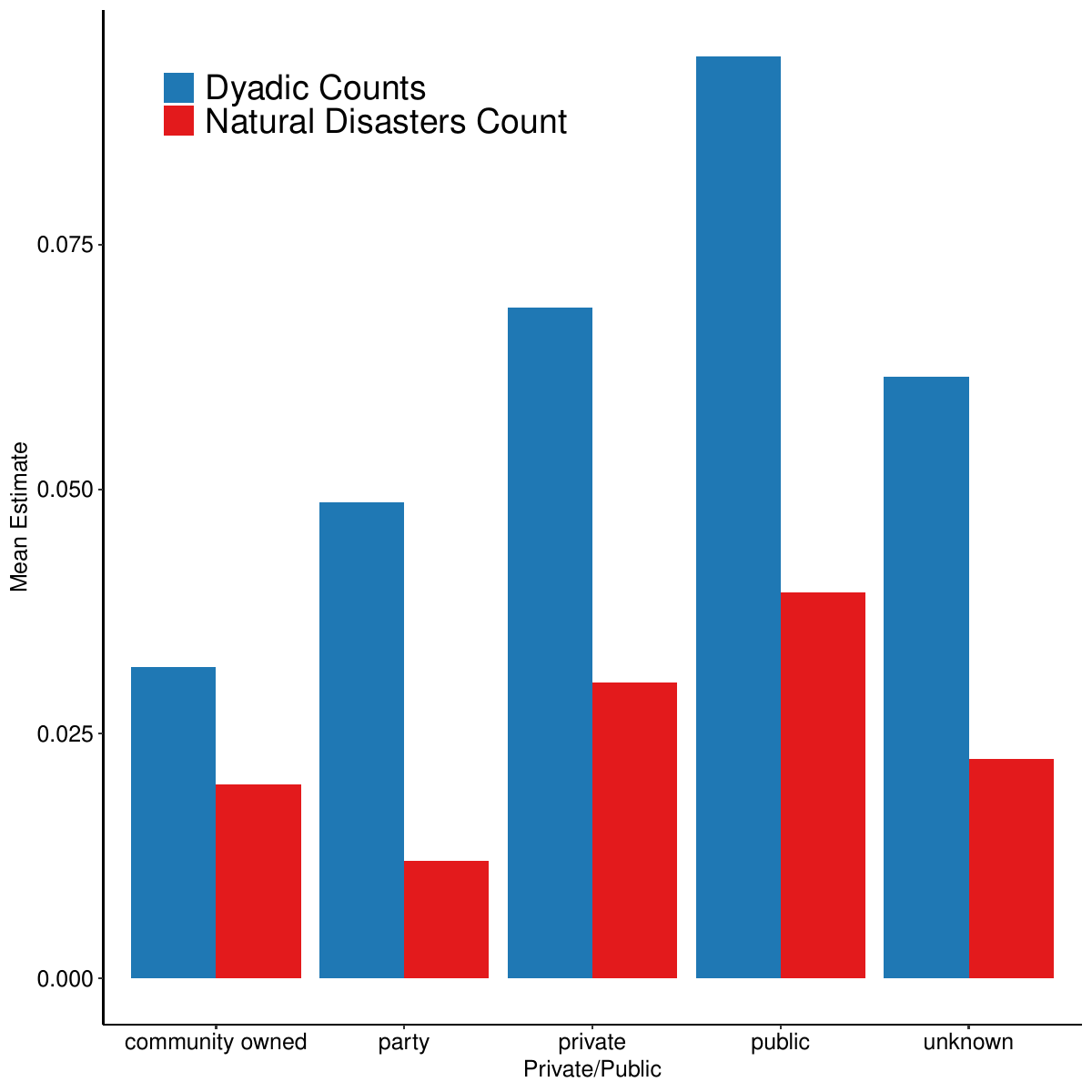}
\label{fig:public_private}
\captionsetup{singlelinecheck=off,font=scriptsize}
\caption*{\textbf{Note}: Figure displays the heterogeneity in the estimated increase in dyadic media coverage following \emdatcount natural disasters that occurred since 2016 by natural disaster type. This figure is constructed taking the estimated increase in media coverage following a natural disaster from the event-level dataset. }
\end{figure}
%%%%%%%%%%%%%%%%%%%%%%%%%%%%%%%%%%%%%%%%%%%%%%%%%%%%%%%%%%%%%%%%%%%%%

%%%%%%%%%%%%%%%%%%%%%%%%%%%%%%%%%%%%%%%%%%%%%%%%%%%%%%%%%%%%%%%%%%%%%
\begin{longtable}{@{} 
    >{\raggedright\arraybackslash}p{1.2cm} 
    >{\centering\arraybackslash}p{2cm} 
    >{\raggedright\arraybackslash}p{10cm} 
    >{\centering\arraybackslash}p{2cm} 
@{}}

    \caption{Tabulation of Sources by Country\label{table:source_list}} \\
    
    % Define alternating row colors starting from the second row
   % \rowcolors{2}{lightgray}{white}
    
    \toprule
    Country &  & URL of the source & Public Sources \\
    \midrule
    \endfirsthead
    
      \midrule
    \multicolumn{4}{r}{\emph{Continued on next page}} \\
    \endfoot
    
    \bottomrule
    \endlastfoot
    
    US & 5 & \href{https://nytimes.com}{nytimes.com}, \href{https://foxnews.com}{foxnews.com}, \href{https://breitbart.com}{breitbart.com}, \href{https://nbcnews.com}{nbcnews.com}, \href{https://foxbusiness.com}{foxbusiness.com} & 0 \\
    GB & 5 & \href{https://bbc.co.uk}{bbc.co.uk}, \href{https://dailymail.co.uk}{dailymail.co.uk}, \href{https://theguardian.com}{theguardian.com}, \href{https://mirror.co.uk}{mirror.co.uk}, \href{https://express.co.uk}{express.co.uk} & 1 \\
    VN & 5 & \href{https://vnexpress.net}{vnexpress.net}, \href{https://vietnamnet.vn}{vietnamnet.vn}, \href{https://dantri.com.vn}{dantri.com.vn}, \href{https://tuoitre.vn}{tuoitre.vn}, \href{https://tienphong.vn}{tienphong.vn} & 3 \\
    NL & 5 & \href{https://nu.nl}{nu.nl}, \href{https://nos.nl}{nos.nl}, \href{https://rtlnieuws.nl}{rtlnieuws.nl}, \href{https://nrc.nl}{nrc.nl}, \href{https://rtvnoord.nl}{rtvnoord.nl} & 2 \\
    IT & 5 & \href{https://corriere.it}{corriere.it}, \href{https://repubblica.it}{repubblica.it}, \href{https://ilfattoquotidiano.it}{ilfattoquotidiano.it}, \href{https://lastampa.it}{lastampa.it}, \href{https://fanpage.it}{fanpage.it} & 0 \\
    DE & 5 & \href{https://bild.de}{bild.de}, \href{https://spiegel.de}{spiegel.de}, \href{https://tagesschau.de}{tagesschau.de}, \href{https://faz.net}{faz.net}, \href{https://sueddeutsche.de}{sueddeutsche.de} & 1 \\
    AU & 3 & \href{https://news.com.au}{news.com.au}, \href{https://abc.net.au}{abc.net.au}, \href{https://9news.com.au}{9news.com.au} & 1 \\
    AR & 5 & \href{https://clarin.com}{clarin.com}, \href{https://lanacion.com.ar}{lanacion.com.ar}, \href{https://ambito.com}{ambito.com}, \href{https://tn.com.ar}{tn.com.ar}, \href{https://cronista.com}{cronista.com} & NA \\
    PL & 5 & \href{https://tvn24.pl}{tvn24.pl}, \href{https://rmf24.pl}{rmf24.pl}, \href{https://wpolityce.pl}{wpolityce.pl}, \href{https://epoznan.pl}{epoznan.pl}, \href{https://gazetaprawna.pl}{gazetaprawna.pl} & 0 \\
    DK & 5 & \href{https://ekstrabladet.dk}{ekstrabladet.dk}, \href{https://bt.dk}{bt.dk}, \href{https://politiken.dk}{politiken.dk}, \href{https://finans.dk}{finans.dk}, \href{https://information.dk}{information.dk} & 0 \\
    FR & 5 & \href{https://lemonde.fr}{lemonde.fr}, \href{https://francetvinfo.fr}{francetvinfo.fr}, \href{https://lefigaro.fr}{lefigaro.fr}, \href{https://bfmtv.com}{bfmtv.com}, \href{https://20minutes.fr}{20minutes.fr} & 1 \\
    ES & 5 & \href{https://elpais.com}{elpais.com}, \href{https://abc.es}{abc.es}, \href{https://eldiario.es}{eldiario.es}, \href{https://elperiodico.com}{elperiodico.com}, \href{https://larazon.es}{larazon.es} & 0 \\
    NO & 5 & \href{https://vg.no}{vg.no}, \href{https://dagbladet.no}{dagbladet.no}, \href{https://aftenposten.no}{aftenposten.no}, \href{https://bt.no}{bt.no}, \href{https://dn.no}{dn.no} & 0 \\
    JP & 5 & \href{https://nikkei.com}{nikkei.com}, \href{https://mainichi.jp}{mainichi.jp}, \href{https://toonippo.co.jp}{toonippo.co.jp}, \href{https://sankei.com}{sankei.com}, \href{https://www3.nhk.or.jp}{www3.nhk.or.jp} & NA \\
    FI & 5 & \href{https://yle.fi}{yle.fi}, \href{https://hs.fi}{hs.fi}, \href{https://uusisuomi.fi}{uusisuomi.fi}, \href{https://kaleva.fi}{kaleva.fi}, \href{https://talouselama.fi}{talouselama.fi} & 1 \\
    AT & 5 & \href{https://orf.at}{orf.at}, \href{https://krone.at}{krone.at}, \href{https://kurier.at}{kurier.at}, \href{https://tt.com}{tt.com}, \href{https://diepresse.com}{diepresse.com} & 1 \\
    SE & 5 & \href{https://aftonbladet.se}{aftonbladet.se}, \href{https://dn.se}{dn.se}, \href{https://di.se}{di.se}, \href{https://svd.se}{svd.se}, \href{https://sverigesradio.se}{sverigesradio.se} & 1 \\
    CA & 5 & \href{https://cbc.ca}{cbc.ca}, \href{https://lapresse.ca}{lapresse.ca}, \href{https://ctvnews.ca}{ctvnews.ca}, \href{https://thestar.com}{thestar.com}, \href{https://vancouversun.com}{vancouversun.com} & 1 \\
    RU & 5 & \href{https://lenta.ru}{lenta.ru}, \href{https://gazeta.ru}{gazeta.ru}, \href{https://vz.ru}{vz.ru}, \href{https://ng.ru}{ng.ru}, \href{https://fedpress.ru}{fedpress.ru} & NA \\
    IN & 5 & \href{https://indianexpress.com}{indianexpress.com}, \href{https://thehindu.com}{thehindu.com}, \href{https://dnaindia.com}{dnaindia.com}, \href{https://firstpost.com}{firstpost.com}, \href{https://newindianexpress.com}{newindianexpress.com} & 0 \\
    CO & 5 & \href{https://eltiempo.com}{eltiempo.com}, \href{https://elcolombiano.com}{elcolombiano.com}, \href{https://caracol.com.co}{caracol.com.co}, \href{https://elheraldo.co}{elheraldo.co}, \href{https://portafolio.co}{portafolio.co} & NA \\
    RO & 5 & \href{https://adevarul.ro}{adevarul.ro}, \href{https://ziare.com}{ziare.com}, \href{https://stiripesurse.ro}{stiripesurse.ro}, \href{https://romaniatv.net}{romaniatv.net}, \href{https://antena3.ro}{antena3.ro} & 0 \\
    EG & 4 & \href{https://youm7.com}{youm7.com}, \href{https://akhbarelyom.com}{akhbarelyom.com}, \href{https://almalnews.com}{almalnews.com}, \href{https://almesryoon.com}{almesryoon.com} & 1 \\
    CH & 5 & \href{https://blick.ch}{blick.ch}, \href{https://20min.ch}{20min.ch}, \href{https://tagesanzeiger.ch}{tagesanzeiger.ch}, \href{https://lematin.ch}{lematin.ch}, \href{https://nzz.ch}{nzz.ch} & 0 \\
    GR & 5 & \href{https://newsit.gr}{newsit.gr}, \href{https://newsbeast.gr}{newsbeast.gr}, \href{https://tvxs.gr}{tvxs.gr}, \href{https://avgi.gr}{avgi.gr}, \href{https://newpost.gr}{newpost.gr} & 0 \\
    NZ & 2 & \href{https://nzherald.co.nz}{nzherald.co.nz}, \href{https://newstalkzb.co.nz}{newstalkzb.co.nz} & 0 \\
    UA & 5 & \href{https://pravda.com.ua}{pravda.com.ua}, \href{https://focus.ua}{focus.ua}, \href{https://gazeta.ua}{gazeta.ua}, \href{https://24tv.ua}{24tv.ua}, \href{https://podrobnosti.ua}{podrobnosti.ua} & 0 \\
    BR & 5 & \href{https://otempo.com.br}{otempo.com.br}, \href{https://em.com.br}{em.com.br}, \href{https://correiobraziliense.com.br}{correiobraziliense.com.br}, \href{https://opovo.com.br}{opovo.com.br}, \href{https://correiodopovo.com.br}{correiodopovo.com.br} & 0 \\
    MY & 2 & \href{https://thestar.com.my}{thestar.com.my}, \href{https://malaysiakini.com}{malaysiakini.com} & 0 \\
    PE & 5 & \href{https://rpp.pe}{rpp.pe}, \href{https://larepublica.pe}{larepublica.pe}, \href{https://americatv.com.pe}{americatv.com.pe}, \href{https://gestion.pe}{gestion.pe}, \href{https://diariocorreo.pe}{diariocorreo.pe} & 0 \\
    IE & 5 & \href{https://independent.ie}{independent.ie}, \href{https://rte.ie}{rte.ie}, \href{https://irishtimes.com}{irishtimes.com}, \href{https://irishexaminer.com}{irishexaminer.com}, \href{https://breakingnews.ie}{breakingnews.ie} & 1 \\
    BE & 5 & \href{https://rtbf.be}{rtbf.be}, \href{https://nieuwsblad.be}{nieuwsblad.be}, \href{https://rtl.be}{rtl.be}, \href{https://standaard.be}{standaard.be}, \href{https://gva.be}{gva.be} & 1 \\
    MX & 5 & \href{https://excelsior.com.mx}{excelsior.com.mx}, \href{https://debate.com.mx}{debate.com.mx}, \href{https://eleconomista.com.mx}{eleconomista.com.mx}, \href{https://lasillarota.com}{lasillarota.com}, \href{https://diario.mx}{diario.mx} & 0 \\
    HN & 4 & \href{https://laprensa.hn}{laprensa.hn}, \href{https://latribuna.hn}{latribuna.hn}, \href{https://tiempo.hn}{tiempo.hn}, \href{https://proceso.hn}{proceso.hn} & 0 \\
    CL & 5 & \href{https://lacuarta.com}{lacuarta.com}, \href{https://cooperativa.cl}{cooperativa.cl}, \href{https://elmostrador.cl}{elmostrador.cl}, \href{https://adnradio.cl}{adnradio.cl}, \href{https://emol.com}{emol.com} & 0 \\
    GT & 2 & \href{https://prensalibre.com}{prensalibre.com}, \href{https://sonora.com.gt}{sonora.com.gt} & 0 \\
    BD & 2 & \href{https://thedailystar.net}{thedailystar.net}, \href{https://bdnews24.com}{bdnews24.com} & 0 \\
    SG & 4 & \href{https://channelnewsasia.com}{channelnewsasia.com}, \href{https://straitstimes.com}{straitstimes.com}, \href{https://tamilmurasu.com.sg}{tamilmurasu.com.sg}, \href{https://eco-business.com}{eco-business.com} & 1 \\
    PK & 5 & \href{https://geo.tv}{geo.tv}, \href{https://dawn.com}{dawn.com}, \href{https://tribune.com.pk}{tribune.com.pk}, \href{https://pakobserver.net}{pakobserver.net}, \href{https://dailytimes.com.pk}{dailytimes.com.pk} & 0 \\
    PT & 5 & \href{https://observador.pt}{observador.pt}, \href{https://rtp.pt}{rtp.pt}, \href{https://dn.pt}{dn.pt}, \href{https://jornaldenegocios.pt}{jornaldenegocios.pt}, \href{https://sabado.pt}{sabado.pt} & 1 \\
    KE & 5 & \href{https://the-star.co.ke}{the-star.co.ke}, \href{https://standardmedia.co.ke}{standardmedia.co.ke}, \href{https://capitalfm.co.ke}{capitalfm.co.ke}, \href{https://businessdailyafrica.com}{businessdailyafrica.com}, \href{https://theeastafrican.co.ke}{theeastafrican.co.ke} & 0 \\
    EC & 5 & \href{https://eluniverso.com}{eluniverso.com}, \href{https://teleamazonas.com}{teleamazonas.com}, \href{https://elcomercio.com}{elcomercio.com}, \href{https://lahora.com.ec}{lahora.com.ec}, \href{https://expreso.ec}{expreso.ec} & 0 \\
    UY & 5 & \href{https://elobservador.com.uy}{elobservador.com.uy}, \href{https://montevideo.com.uy}{montevideo.com.uy}, \href{https://teledoce.com}{teledoce.com}, \href{https://diarioelpueblo.com.uy}{diarioelpueblo.com.uy}, \href{https://carasycaretas.com.uy}{carasycaretas.com.uy} & 0 \\
    DO & 5 & \href{https://diariolibre.com}{diariolibre.com}, \href{https://listindiario.com}{listindiario.com}, \href{https://almomento.net}{almomento.net}, \href{https://elcaribe.com.do}{elcaribe.com.do}, \href{https://hoy.com.do}{hoy.com.do} & 0 \\
    IL & 5 & \href{https://timesofisrael.com}{timesofisrael.com}, \href{https://ynetnews.com}{ynetnews.com}, \href{https://arab48.com}{arab48.com}, \href{https://sonara.net}{sonara.net}, \href{https://ynet.co.il}{ynet.co.il} & 0 \\
    MA & 5 & \href{https://lematin.ma}{lematin.ma}, \href{https://yabiladi.com}{yabiladi.com}, \href{https://bladi.net}{bladi.net}, \href{https://telquel.ma}{telquel.ma}, \href{https://lavieeco.com}{lavieeco.com} & 0 \\
    TR & 5 & \href{https://gazetevatan.com}{gazetevatan.com}, \href{https://haberler.com}{haberler.com}, \href{https://mynet.com}{mynet.com}, \href{https://sabah.com.tr}{sabah.com.tr}, \href{https://haber7.com}{haber7.com} & 0 \\
    NG & 5 & \href{https://saharareporters.com}{saharareporters.com}, \href{https://dailypost.ng}{dailypost.ng}, \href{https://guardian.ng}{guardian.ng}, \href{https://pulse.ng}{pulse.ng}, \href{https://sunnewsonline.com}{sunnewsonline.com} & 0 \\
    PA & 4 & \href{https://laestrella.com.pa}{laestrella.com.pa}, \href{https://panamaamerica.com.pa}{panamaamerica.com.pa}, \href{https://tvn-2.com}{tvn-2.com}, \href{https://rpctv.com}{rpctv.com} & 0 \\
    VE & 5 & \href{https://lapatilla.com}{lapatilla.com}, \href{https://ultimasnoticias.com.ve}{ultimasnoticias.com.ve}, \href{https://globovision.com}{globovision.com}, \href{https://elimpulso.com}{elimpulso.com}, \href{https://unionradio.net}{unionradio.net} & 0 \\
    BO & 4 & \href{https://eldeber.com.bo}{eldeber.com.bo}, \href{https://lostiempos.com}{lostiempos.com}, \href{https://erbol.com.bo}{erbol.com.bo}, \href{https://hoybolivia.com}{hoybolivia.com} & 0 \\
    AE & 4 & \href{https://gulfnews.com}{gulfnews.com}, \href{https://emaratalyoum.com}{emaratalyoum.com}, \href{https://khaleejtimes.com}{khaleejtimes.com}, \href{https://gulftoday.ae}{gulftoday.ae} & NA \\
    KZ & 5 & \href{https://inform.kz}{inform.kz}, \href{https://kt.kz}{kt.kz}, \href{https://kp.kz}{kp.kz}, \href{https://zonakz.net}{zonakz.net}, \href{https://azh.kz}{azh.kz} & 1 \\
    HK & 3 & \href{https://scmp.com}{scmp.com}, \href{https://takungpao.com.hk}{takungpao.com.hk}, \href{https://orientaldaily.on.cc}{orientaldaily.on.cc} & 0 \\
    SA & 5 & \href{https://aawsat.com}{aawsat.com}, \href{https://slaati.com}{slaati.com}, \href{https://okaz.com.sa}{okaz.com.sa}, \href{https://almowaten.net}{almowaten.net}, \href{https://al-madina.com}{al-madina.com} & 0 \\
    DZ & 5 & \href{https://echoroukonline.com}{echoroukonline.com}, \href{https://elkhabar.com}{elkhabar.com}, \href{https://entv.dz}{entv.dz}, \href{https://lequotidien-oran.com}{lequotidien-oran.com}, \href{https://lnr-dz.com}{lnr-dz.com} & 2 \\
    ID & 5 & \href{https://thejakartapost.com}{thejakartapost.com}, \href{https://mediaindonesia.com}{mediaindonesia.com}, \href{https://regional.kompas.com}{regional.kompas.com}, \href{https://kabar24.bisnis.com}{kabar24.bisnis.com}, \href{https://inet.detik.com}{inet.detik.com} & 0 \\
    CR & 4 & \href{https://larepublica.net}{larepublica.net}, \href{https://elpais.cr}{elpais.cr}, \href{https://monumental.co.cr}{monumental.co.cr}, \href{https://thecostaricanews.com}{thecostaricanews.com} & NA \\
    LB & 5 & \href{https://almanar.com.lb}{almanar.com.lb}, \href{https://aljadeed.tv}{aljadeed.tv}, \href{https://tayyar.org}{tayyar.org}, \href{https://annahar.com}{annahar.com}, \href{https://almayadeen.net}{almayadeen.net} & 0 \\
    ZW & 5 & \href{https://herald.co.zw}{herald.co.zw}, \href{https://newsday.co.zw}{newsday.co.zw}, \href{https://bulawayo24.com}{bulawayo24.com}, \href{https://thezimbabwemail.com}{thezimbabwemail.com}, \href{https://sundaymail.co.zw}{sundaymail.co.zw} & 2 \\
    JM & 3 & \href{https://jamaicaobserver.com}{jamaicaobserver.com}, \href{https://jamaica-star.com}{jamaica-star.com}, \href{https://jamaica-gleaner.com}{jamaica-gleaner.com} & 0 \\
    LK & 3 & \href{https://dailymirror.lk}{dailymirror.lk}, \href{https://colombopage.com}{colombopage.com}, \href{https://island.lk}{island.lk} & 0 \\
    KW & 3 & \href{https://alanba.com.kw}{alanba.com.kw}, \href{https://aljarida.com}{aljarida.com}, \href{https://arabtimesonline.com}{arabtimesonline.com} & 0 \\
    PH & 5 & \href{https://tribune.net.ph}{tribune.net.ph}, \href{https://news.abs-cbn.com}{news.abs-cbn.com}, \href{https://manilatimes.net}{manilatimes.net}, \href{https://sunstar.com.ph}{sunstar.com.ph}, \href{https://bworldonline.com}{bworldonline.com} & 0 \\
    QA & 4 & \href{https://aljazeera.net}{aljazeera.net}, \href{https://al-sharq.com}{al-sharq.com}, \href{https://alarab.qa}{alarab.qa}, \href{https://thepeninsulaqatar.com}{thepeninsulaqatar.com} & 1 \\
    JO & 5 & \href{https://addustour.com}{addustour.com}, \href{https://alghad.com}{alghad.com}, \href{https://alrai.com}{alrai.com}, \href{https://jordantimes.com}{jordantimes.com}, \href{https://en.ammonnews.net}{en.ammonnews.net} & 0 \\
    UG & 2 & \href{https://observer.ug}{observer.ug}, \href{https://monitor.co.ug}{monitor.co.ug} & 0 \\
    TT & 3 & \href{https://trinidadexpress.com}{trinidadexpress.com}, \href{https://newsday.co.tt}{newsday.co.tt}, \href{https://tv6tnt.com}{tv6tnt.com} & 0 \\
    LU & 5 & \href{https://lessentiel.lu}{lessentiel.lu}, \href{https://wort.lu}{wort.lu}, \href{https://rtl.lu}{rtl.lu}, \href{https://tageblatt.lu}{tageblatt.lu}, \href{https://lequotidien.lu}{lequotidien.lu} & NA \\
    TN & 2 & \href{https://lapresse.tn}{lapresse.tn}, \href{https://leconomistemaghrebin.com}{leconomistemaghrebin.com} & 1 \\
    AM & 2 & \href{https://panarmenian.net}{panarmenian.net}, \href{https://golosarmenii.am}{golosarmenii.am} & NA \\
    CG & 2 & \href{https://adiac-congo.com}{adiac-congo.com}, \href{https://journaldebrazza.com}{journaldebrazza.com} & 0 \\
    PS & 1 & \href{https://alquds.com}{alquds.com} & 0 \\
    TZ & 3 & \href{https://thecitizen.co.tz}{thecitizen.co.tz}, \href{https://habarileo.co.tz}{habarileo.co.tz}, \href{https://mtanzania.co.tz}{mtanzania.co.tz} & 1 \\
    HT & 1 & \href{https://haitilibre.com}{haitilibre.com} & NA \\
    NE & 1 & \href{https://actuniger.com}{actuniger.com} & 0 \\
    IR & 5 & \href{https://iribnews.ir}{iribnews.ir}, \href{https://aryanews.com}{aryanews.com}, \href{https://khabaronline.ir}{khabaronline.ir}, \href{https://tabnak.ir}{tabnak.ir}, \href{https://fardanews.com}{fardanews.com} & NA \\
    CN & 5 & \href{https://finance.sina.com.cn}{finance.sina.com.cn}, \href{https://news.dayoo.com}{news.dayoo.com}, \href{https://china.qianlong.com}{china.qianlong.com}, \href{https://world.people.com.cn}{world.people.com.cn}, \href{https://news.southcn.com}{news.southcn.com} & 3 \\
    RS & 5 & \href{https://blic.rs}{blic.rs}, \href{https://b92.net}{b92.net}, \href{https://novosti.rs}{novosti.rs}, \href{https://rts.rs}{rts.rs}, \href{https://politika.rs}{politika.rs} & NA \\
    AZ & 2 & \href{https://news.day.az}{news.day.az}, \href{https://azernews.az}{azernews.az} & 0 \\
    BG & 5 & \href{https://24chasa.bg}{24chasa.bg}, \href{https://focus-news.net}{focus-news.net}, \href{https://fakti.bg}{fakti.bg}, \href{https://bnr.bg}{bnr.bg}, \href{https://dnesplus.bg}{dnesplus.bg} & 1 \\
    HU & 5 & \href{https://origo.hu}{origo.hu}, \href{https://index.hu}{index.hu}, \href{https://blikk.hu}{blikk.hu}, \href{https://hvg.hu}{hvg.hu}, \href{https://hirado.hu}{hirado.hu} & NA \\
    SK & 5 & \href{https://teraz.sk}{teraz.sk}, \href{https://topky.sk}{topky.sk}, \href{https://cas.sk}{cas.sk}, \href{https://24hod.sk}{24hod.sk}, \href{https://aktuality.sk}{aktuality.sk} & 1 \\
    HR & 5 & \href{https://vecernji.hr}{vecernji.hr}, \href{https://tportal.hr}{tportal.hr}, \href{https://net.hr}{net.hr}, \href{https://nacional.hr}{nacional.hr}, \href{https://dnevnik.hr}{dnevnik.hr} & 0 \\
    TH & 4 & \href{https://matichon.co.th}{matichon.co.th}, \href{https://thairath.co.th}{thairath.co.th}, \href{https://bangkokpost.com}{bangkokpost.com}, \href{https://thaipost.net}{thaipost.net} & 0 \\
    LT & 5 & \href{https://15min.lt}{15min.lt}, \href{https://lrt.lt}{lrt.lt}, \href{https://tv3.lt}{tv3.lt}, \href{https://delfi.lt}{delfi.lt}, \href{https://kauno.diena.lt}{kauno.diena.lt} & 1 \\
    BY & 4 & \href{https://charter97.org}{charter97.org}, \href{https://afn.by}{afn.by}, \href{https://vb.by}{vb.by}, \href{https://bdg.by}{bdg.by} & 0 \\
    KR & 2 & \href{https://world.kbs.co.kr}{world.kbs.co.kr}, \href{https://english.chosun.com}{english.chosun.com} & 1 \\
    SI & 5 & \href{https://siol.net}{siol.net}, \href{https://rtvslo.si}{rtvslo.si}, \href{https://delo.si}{delo.si}, \href{https://slovenskenovice.si}{slovenskenovice.si}, \href{https://vecer.com}{vecer.com} & 1 \\
    CZ & 5 & \href{https://parlamentnilisty.cz}{parlamentnilisty.cz}, \href{https://lidovky.cz}{lidovky.cz}, \href{https://novinky.cz}{novinky.cz}, \href{https://tn.nova.cz}{tn.nova.cz}, \href{https://denik.cz}{denik.cz} & 0 \\
    BA & 2 & \href{https://avaz.ba}{avaz.ba}, \href{https://glassrpske.com}{glassrpske.com} & 0 \\
    CY & 5 & \href{https://dialogos.com.cy}{dialogos.com.cy}, \href{https://kathimerini.com.cy}{kathimerini.com.cy}, \href{https://havadiskibris.com}{havadiskibris.com}, \href{https://halkinsesikibris.com}{halkinsesikibris.com}, \href{https://yeniduzen.com}{yeniduzen.com} & NA \\
    SD & 3 & \href{https://alnilin.com}{alnilin.com}, \href{https://sudanile.com}{sudanile.com}, \href{https://dabangasudan.org}{dabangasudan.org} & 0 \\
    LV & 5 & \href{https://delfi.lv}{delfi.lv}, \href{https://diena.lv}{diena.lv}, \href{https://db.lv}{db.lv}, \href{https://ventasbalss.lv}{ventasbalss.lv}, \href{https://baltictimes.com}{baltictimes.com} & 0 \\
    PY & 5 & \href{https://lanacion.com.py}{lanacion.com.py}, \href{https://hoy.com.py}{hoy.com.py}, \href{https://nanduti.com.py}{nanduti.com.py}, \href{https://paraguay.com}{paraguay.com}, \href{https://5dias.com.py}{5dias.com.py} & 0 \\
    NP & 1 & \href{https://gorkhapatraonline.com}{gorkhapatraonline.com} & 1 \\
    ZA & 4 & \href{https://citizen.co.za}{citizen.co.za}, \href{https://news24.com}{news24.com}, \href{https://ewn.co.za}{ewn.co.za}, \href{https://mg.co.za}{mg.co.za} & 0 \\
    EE & 5 & \href{https://ohtuleht.ee}{ohtuleht.ee}, \href{https://delfi.ee}{delfi.ee}, \href{https://aripaev.ee}{aripaev.ee}, \href{https://postimees.ee}{postimees.ee}, \href{https://news.err.ee}{news.err.ee} & 1 \\
    BH & 2 & \href{https://tradearabia.com}{tradearabia.com}, \href{https://gdnonline.com}{gdnonline.com} & 0 \\
    ML & 4 & \href{https://news.abamako.com}{news.abamako.com}, \href{https://maliweb.net}{maliweb.net}, \href{https://malijet.com}{malijet.com}, \href{https://journaldumali.com}{journaldumali.com} & 0 \\
    PR & 2 & \href{https://elvocero.com}{elvocero.com}, \href{https://elexpresso.com}{elexpresso.com} & 0 \\
    TW & 1 & \href{https://taipeitimes.com}{taipeitimes.com} & 0 \\
    CU & 5 & \href{https://radiohc.cu}{radiohc.cu}, \href{https://cubadebate.cu}{cubadebate.cu}, \href{https://radiorebelde.cu}{radiorebelde.cu}, \href{https://sierramaestra.cu}{sierramaestra.cu}, \href{https://escambray.cu}{escambray.cu} & NA \\
    KH & 2 & \href{https://akp.gov.kh}{akp.gov.kh}, \href{https://phnompenhpost.com}{phnompenhpost.com} & 0 \\
    OM & 3 & \href{https://shabiba.com}{shabiba.com}, \href{https://timesofoman.com}{timesofoman.com}, \href{https://omanobserver.om}{omanobserver.om} & 1 \\
    CM & 1 & \href{https://journalducameroun.com}{journalducameroun.com} & 0 \\
    IQ & 5 & \href{https://azzaman.com}{azzaman.com}, \href{https://alsabahpress.com}{alsabahpress.com}, \href{https://aliraqnews.com}{aliraqnews.com}, \href{https://albasrah.net}{albasrah.net}, \href{https://ekurd.net}{ekurd.net} & 1 \\
    YE & 3 & \href{https://26sep.net}{26sep.net}, \href{https://almotamar.net}{almotamar.net}, \href{https://alwahdawi.net}{alwahdawi.net} & 1 \\
    TD & 1 & \href{https://alwihdainfo.com}{alwihdainfo.com} & 0 \\
    SN & 2 & \href{https://rewmi.com}{rewmi.com}, \href{https://news.adakar.com}{news.adakar.com} & 0 \\
    SV & 2 & \href{https://diariocolatino.com}{diariocolatino.com}, \href{https://diario1.com}{diario1.com} & 0 \\
    GN & 5 & \href{https://aminata.com}{aminata.com}, \href{https://guineenews.org}{guineenews.org}, \href{https://africaguinee.com}{africaguinee.com}, \href{https://kababachir.com}{kababachir.com}, \href{https://guineemining.info}{guineemining.info} & 0 \\
    CI & 1 & \href{https://fratmat.info}{fratmat.info} & 1 \\
    BF & 2 & \href{https://lefaso.net}{lefaso.net}, \href{https://news.aouaga.com}{news.aouaga.com} & 0 \\
    CD & 1 & \href{https://radiookapi.net}{radiookapi.net} & 1 \\
    AF & 1 & \href{https://avapress.com}{avapress.com} & NA \\
    MM & 1 & \href{https://irrawaddy.com}{irrawaddy.com} & 0 \\
    GE & 2 & \href{https://civil.ge}{civil.ge}, \href{https://messenger.com.ge}{messenger.com.ge} & 0 \\
    MD & 1 & \href{https://373news.com}{373news.com} & 0 \\
    MG & 1 & \href{https://midi-madagasikara.mg}{midi-madagasikara.mg} & 0 \\
    BW & 2 & \href{https://mmegi.bw}{mmegi.bw}, \href{https://sundaystandard.info}{sundaystandard.info} & 0 \\
    GM & 1 & \href{https://thepoint.gm}{thepoint.gm} & 0 \\
    SS & 1 & \href{https://radiotamazuj.org}{radiotamazuj.org} & 0 \\
    
\end{longtable}
%%%%%%%%%%%%%%%%%%%%%%%%%%%%%%%%%%%%%%%%%%%%%%%%%%%%%%%%%%%%%%%%%%%%%%

\end{document}